\RequirePackage{fix-cm}
\documentclass[smallcondensed]{svjour3}     
\smartqed  
\usepackage{graphicx}
\usepackage{amssymb}
\usepackage{latexsym}
\usepackage{color}
\usepackage{natbib}
\newcommand{\arcsec}{$^{\prime\prime}$}
\newcommand{\msun}{$M_{\odot}$}
\newcommand{\ct}{\citealt}

\begin{document}

\title{Formation and Evolution of Disks around Young Stellar Objects
}


\author{Bo Zhao \and Kengo Tomida \and Patrick Hennebelle \and John J. Tobin \and Ana\"{e}lle Maury \and Tomoya Hirota \and \'{A}lvaro S\'{a}nchez-Monge \and Rolf Kuiper \and Anna Rosen\and Asmita Bhandare \and Marco Padovani  \and Yueh-Ning Lee 
}


\institute{B. Zhao \at
              Max-Planck-Institute for Extraterrestrial Physics, Giessenbachstr. 1, 85748 Garching, Germany \\
              \email{bo.zhao@mpe.mpg.de}           
           \and
           K. Tomida \at 
              Astronomical Institute, Tohoku University, 6-3, Aramaki, Aoba-ku, Sendai, Miyagi 980-8578, Japan
           \and
           P. Hennebelle, A. Maury \at
              AIM/CEA/CNRS, Universit\'{e} Paris-Saclay, Universit\'{e} Paris Diderot, Sorbonne Paris Cit\'{e}, 91191 Gif sur Yvette, France
           \and
           J. J. Tobin \at
              National Radio Astronomy Observatory 520 Edgemont Road, Charlottesville, VA 22901, USA
           \and
           T. Hirota \at
              National Astronomical Observatory of Japan, Osawa, Mitaka, Tokyo 181-8588, Japan
           \and
           \'{A}. S\'{a}nchez-Monge \at
              I. Physikalisches Institut, Universit\"{a}t zu K\"{o}ln, Z\"{u}lpicher Str. 77, 50937 K\"{o}ln, Germany
           \and
           R. Kuiper \at
              Institute of Astronomy and Astrophysics, University of T\"{u}bingen, Auf der Morgenstelle 10, 72076, T\"{u}bingen, Germany
           \and
           A. Rosen \at
              Center for Astrophysics, Harvard \& Smithsonian, 60 Garden St., Cambridge, MA 02138, USA
           \and
           A. Bhandare \at
              Max Planck Institute for Astronomy, K\"{o}nigstuhl 17, 69117 Heidelberg, Germany
           \and
           M. Padovani \at
              INAF-Osservatorio Astrofisico di Arcetri, Largo E. Fermi, 5 - 50125 Firenze, Italy
           \and
           Y.-N. Lee \at
              Department of Earth Sciences, National Taiwan Normal University, 88, Sec. 4, Ting-Chou Road, Taipei 11677, Taiwan
}

\date{Received: Feb 2020 / Accepted: 2020}

\maketitle

\begin{abstract}
Recent observations have suggested that circumstellar disks may commonly form around young stellar objects. Although the formation of circumstellar disks can be a natural result of the conservation of angular momentum in the parent cloud, theoretical studies instead show disk formation to be difficult from dense molecular cores magnetized to a realistic level, owing to efficient magnetic braking that transports a large fraction of the angular momentum away from the circumstellar region. We review recent progress in the formation and early evolution of disks around young stellar objects of both low-mass and high-mass, with an emphasis on mechanisms that may bridge the gap between observation and theory, including non-ideal MHD effects and asymmetric perturbations in the collapsing core (e.g., magnetic field misalignment and turbulence). We also address the associated processes of outflow launching and the formation of multiple systems, and discuss possible implications in properties of protoplanetary disks.
\end{abstract}

\section{Introduction}
\label{Chap.Intro}

In the process of star formation, circumstellar disks are common structures that form as a result of the conservation of angular momentum of the parent molecular cloud. The advancement of instruments in recent decades such as the Atacama Large Millimeter Array (ALMA) have enabled observations to resolve the small scale disk structures around young stellar objects in nearby star forming regions, many of which show evidence of Keplerian rotation. Ring-like dust sub-structures have also been discovered in disks around low mass protostars, which may be the imprint of early dust evolution and may facilitate the subsequent formation of planets. Therefore, circumstellar disks plays a crucial role in star and planet formation.

The formation of the initial disk structure at the earliest stages of star formation is closely related to the so-called first hydrostatic core (FHSC; \ct{Larson1969}), which forms prior to the protostar as a result of the isothermal collapse of the parent molecular cloud core. The FHSC is predominantly supported by thermal pressure and traps the radiation (optically thick) with the central temperature rapidly reaching $\approx$ 2000~k, around which the hydrogen molecules H$_2$ start to dissociate. The endothermic process triggers the second collapse that eventually forms the protostellar core. Once the temperature at the center of the core rises to $10^6$~K, deuterium burning is initiated. 
During the process, the rotationally supported disks (RSDs) grow from outside the FHSC as collapse continues to bring angular momentum into the central region.

There are many well-known problems in star and disk formation, among which the angular momentum problem, magnetic flux problem, and magnetic braking ``catastrophe" have been the main topics of theoretical and numerical studies in the past few decades. 
\begin{enumerate}
\item The angular momentum problem refers to the fact that young stellar object would rotate much faster than its break-up speed, if the angular momentum of the molecular cloud core were conserved during the gravitational collapse. Although the cloud rotation inferred from observations tend to be slow, with typical ratios of rotational energy to gravitational potential energy $\beta \approx$ 0.02 (\ct{Goodman+1993,Caselli+2002b}), the specific angular momentum still needs to decrease 4--5 orders of magnitude during the collapse. Possible solutions includes forming multiple stellar systems, or circumstellar disks. However, without additional mechanisms to transport away the angular momentum, the resulting disks are as large as a few 10$^2$~au, which are larger than the typical disk sizes observed around young stellar objects (e.g., \ct{Maury+2010}). A plausible candidate for resolving the issue is magnetic braking. 
\item The magnetic flux problem refers to the issue that if the magnetic flux threading the collapsing dense core were to be dragged into a young stellar object, the stellar magnetic field strength would be tens of millions of Gauss, more than three orders of magnitude higher than the typical values (kilo-Gauss range; \ct{Johns-Krull2009}) observed in young stars. Therefore, most of the magnetic flux has to be lost in the course of forming stars. 
\item Magnetic braking (\ct{MestelSpitzer1956}), although being a plausible solution for the angular momentum problem, can instead transport away too much angular momentum so that disk formation is completely suppressed, given a relatively strong coupling between the magnetic field and the neutral matter. This is so-called magnetic braking ``catastrophe". As dense molecular cores are observed to be relatively strongly magnetized, with a dimensionless mass-to-flux ratio of a few (\ct{TrolandCrutcher2008}), the magnetic braking ``catastrophe" has become a major obstacle for theories to explain disk observations.
\item Massive star formation (O-type and early B-type, with masses above 10~$M_\odot$) also faces the long-standing radiation pressure problem, i.e., the strong stellar radiation feedback of an early, still-accreting high-mass star can halt and stop its formation process. 
The bipolar cavity associated with the circumstellar disk may allow the escape of the strong stellar radiation.
\end{enumerate}

The transport of angular momentum by magnetic field is closely related to the outflow activities observed in many young stellar objects. Two types of outflows are becoming more recognized in observations as two distinctive entities: the first type is wide-opening slow (a few to $10~{\rm km \,~s^{-1}}$) molecular outflows, which are typically observed with molecular lines like CO; the second is well-collimated fast ($\gtrsim 100~{\rm km\, ~s^{-1}}$) jets, which are typically observed as optical jets. 
Recent observations have also discovered rotation in the bipolar molecular outflow, which provides robust test of the outflow launching mechanism. 

Driven by the rich and diverse disk structures resolved by the high-resolution interferometry at submillimeter to millimeter wavelengths, theoretical work starts to include more physics and provides more realistic conditions to study the formation and early evolution of disks around protostars. The main mechanisms considered in the literature include non-ideal MHD and microphysics, misalignment of magnetic fields, turbulence, and radiation transfer. Currently, evolving the disk for a longer time, i.e. until the Class 0/I phase, is often hindered due to time step restrictions. The next steps in disk formation and evolution studies would involve linking the theoretical models to the recent wealth of observational data, in order to strengthen our understanding of the star-disk connection. 

In this article, we review the recent discoveries in disk observations and the recent theoretical efforts. We start with an analytical description of the fundamental physical process involved in disk formation in general (\S~\ref{Chap.Anly}). The low mass case is presented first in \S~\ref{Chap.DiskLowMass}, covering both observational and numerical developments. The high mass case is presented in a similar fashion in \S~\ref{Chap.DiskHighMass}. We review separately recent studies on gravitational instability in \S~\ref{Chap.GIFrag} and on outflow and jet in \S~\ref{Chap.Outflow}. Finally, we discuss existing observational caveats and numerical limitations in \S~\ref{Chap.Discuss}, along with advancements in synthetic observations that connect numerical models with observations. We summarize the review in \S~\ref{Chap.Summary}.

\section{Fundamental Processes of Disk Formation: Analytical Considerations}
\label{Chap.Anly}

As the physical state from which molecular clouds start collapsing is expected, on one 
hand to be determinant but on the other hand difficult to determine, and even more fundamentally 
subject to a broad distribution, the task of obtaining the protoplanetary disk distributions seems
at first sight complicated. Since the cloud evolution is governed by gravity, thermal pressure, rotation, magnetic field and turbulence, the space of parameters that needs to be explored is indeed rather large. To understand how and in which conditions disks form, we first start with analytical considerations and demonstrate the importance of the physical parameters involved.

\subsection{Axisymmetrical and Rotating Unmagnetized Cloud}

The simplest configuration is obviously a rotating and axisymmetric cloud without 
magnetic field. In this case angular momentum is conserved. 
Let us consider a star of mass $M_*$ and a fluid particle whose specific angular momentum is 
$j$. The centrifugal radius, $r_d$, is obtained when centrifugal and gravitational forces are equal, 
which leads to
\begin{eqnarray}
r_d = {j^2 \over GM_*}.
\label{centrifug}
\end{eqnarray}
From this expression it is clear that the disk formation in a collapsing cloud, depends on 
the initial specific angular momentum distribution. To illustrate this more clearly
let us consider a cloud in solid body rotation, whose density profile is initially proportional to 
$1/r^2$ (e.g., \ct{Shu1977}). In this case, the gas mass enclosed initially in a sphere of 
radius $R_0$ is $M(R_0) \propto R_0$ while the specific angular momentum 
$j(R_0) = R_0^2 \Omega$. This leads to $j(R_0) \propto M(R_0)^2$. As accretion proceeds, 
the shells of increasing initial radius,$R_0$, reach the star-disk system. The centrifugal radius 
for a fluid particle initially located in $R_0$ is therefore given by 
$r_d = j(R_0)^2 / (G M(R_0)) \propto M(R_0)^3$ (\ct{Terebey+1984}).

On the other hand, for a cloud in solid body rotation for which the density is initially  uniform, we have $M(R_0) \propto R_0^3$ and $j \propto M(R_0)^{2/3}$ leading to $r_d \propto M(R_0)^{1/3}$. As the mass delivered within the cloud is typically of the order of a few $C_s^3/ (G) t$, we see that while in the first case, the disk grows like $t^{3}$, i.e. extremely slowly (since $t \simeq 0$ initially), it grows considerably faster in the case of a uniform density cloud.

These two examples illustrate that in the hydrodynamical case, the disk radius depends very sensitively on the angular momentum distribution. It is enlightening to infer more quantitative estimate. Let us consider a spherical cloud of density $\rho_0$ in solid body rotation at a rate $\Omega_0$.
When a particle initially at radius $R_0$ reaches centrifugal equilibrium into the disk, we get
\begin{eqnarray}
r_\mathrm{d,hydro} \simeq {\Omega_0^2 R_0^4 \over 4 \pi / 3 \rho_0 R_0^3 G } = 3 \beta R_0  = 106 \, {\rm AU} \, 
{\beta  \over 0.02 }
\, \left( {M \over 0.1 M _\odot}\right) ^{1/3}  \left( {\rho_0 \over 10^{-18} {\rm g \, cm}^{-3} }\right)^{-1/3},
\end{eqnarray}
where $\beta =R_0^3 \Omega_0^2/ 3GM$ is the ratio of rotational over gravitational energy. As expected, the radius  depends quadratically on the initial rotation rate. As cores have a typical $\beta \simeq 0.02$ (\ct{Goodman+1993,Belloche2013}), pure hydrodynamical disks should have a radius on the order of 100 au. 

\subsection{Impact of Magnetic Braking}

In the presence of magnetic field, the magnetic tension can generate a torque and angular momentum can be exchanged between the various parts of the clouds. To understand the essence of the magnetic braking and its influence on disks, it is useful to start with simple estimates. 

Two timescales are particularly relevant namely, the  magnetic braking  and the rotation time. 
They are given by
\begin{eqnarray}
\label{eq_br}
\tau _{\rm br} &\simeq& { \rho v_\phi 4 \pi h \over B_z B_\phi}, \\
\tau _{\rm rot} &\simeq& { 2 \pi r \over v_ \phi}~,
\label{eq_rot}
\end{eqnarray}
in which $h$ is the scale height of the disk. The analysis consists in estimating these two times at the edge of the disk. If the braking time is shorter than the rotation time, it means that before a gas particle has accomplished a full rotation, it has already lost most of its momentum. To proceed, we need to know the toroidal field $B_\phi$. This latter is generated by the twisting of the field lines and in the ideal MHD limit, it continuously grows. In a time $\tau _{\rm br}$, we get
\begin{eqnarray}
B_\phi \simeq \tau _{\rm br} { B_z v_\phi \over h},
\label{bphi}
\end{eqnarray}
which in combination with Eq.~\ref{eq_br} leads to 
\begin{eqnarray}
\label{eq_br2}
{ \tau _{\rm br} \over \tau _{\rm rot} } &\simeq& {  v_\phi \sqrt{ 4 \pi \rho  h^2} \over 2 \pi B_z r_d }, 
\end{eqnarray}

Then, we assume that the gas in the neighbourhood of the disk outer part has 
a Keplerian velocity (given a disk is formed):
\begin{eqnarray}
v_\phi &\simeq& \sqrt{ G (M_* + M_{\rm  d}) \over r}, 
\label{vphi}
\end{eqnarray}
where $M_{\rm d}$ is the mass of the disk, $M_*$ the mass of the central star. The mass of the disk itself is about $M_{\rm d} \simeq   \pi r_d^2 \rho h $. With this expression Eqs.~\ref{vphi} and~\ref{eq_br2} leads to
\begin{eqnarray}
\label{eq_br3}
{ \tau _{\rm br} \over \tau _{\rm rot} } &\simeq& \left( { M_* + M_ {\rm d} \over M_{\rm d} } \times { h \over r_d } \right)^{1/2} 
{ G^{1/2} \Sigma \over 2  B_z  } \simeq \left(  { M_* \over M_{\rm d} }
{ h \over r_d } \right)^{1/2}   { \lambda _{eff} \over 4 \pi   }, 
\end{eqnarray}
where $\Sigma=2 h \rho$, is the disk column density and $\lambda_{eff} =  2 \pi G^{1/2} \Sigma / B$ is the mass to flux ratio calculated through the disk. This latter is expected to be lower than the core mass to flux ratio as long as ideal MHD holds. Equation~\ref{eq_br3} clearly leads to the conclusion that even with modest magnetic intensities corresponding to values as high as few times $ 4 \pi$, the magnetic braking is acting on a timescale shorter than the rotation time. As time goes on, that is to say when the central star grows in mass, $M_* / M_{\rm d}$  becomes larger and it should be easier to form a disk. Nevertheless, Eq.~\ref{eq_br3} clearly shows that even weak field, should make a strong difference by suppressing the formation of RSDs. This process is known as the catastrophic magnetic braking. \citet{Galli+2006} have analyzed in detail this process using an analytical expansion of self-similar solutions applied to collapsing dense cores. They also found that magnetic field suppresses disk formation even for modest intensities.

Finally, we note that the analysis mostly applies to the disk region. In fact, toroidal components of the magnetic field can already develop in the inner envelope ($\lesssim$ a few 10$^2$~au), thus magnetic braking is also important in the inner pseudo-disk region outside the disk, which determines the angular momentum influx into the disk. Therefore, physical processes retaining gas angular momentum and reducing magnetic braking efficiency outside (rather than inside) the disk is the key when considering the formation of disks. We will elaborate on the non-ideal MHD effects as well as non-axisymmetric perturbations of the parent molecular cloud in \S~\ref{S.IC_impact}. For the later evolution of protoplanetary disks after the depletion of the infalling envelope, one should investigate the physical processes in the disk itself.

\section{Disk Formation in Low Mass Stars}
\label{Chap.DiskLowMass}

\subsection{Recent observations of embedded disks}
\label{S.ObsLowMass}

We now review the recent observational discoveries of disks around solar-type protostars. 
The high resolution and sensitivity of ALMA is not only enabling larger samples of protostellar dust disk to be characterized, but it is also allowing molecular line observations toward protostellar systems with unprecedented sensitivity. This enables the kinematics of the protostellar envelope down to the disk to be measured with both high sensitivity and sub-arcsecond resolution.
As a result, observations start to reveal the detailed  structures of the infalling-rotating envelope, identify the Keplerian rotation, asymmetric and small scale structures within disks, which pose challenges to existing theories of disk formation.


\subsubsection{Observational Characterization}

The disks toward protostars can be observed in the submm/mm from the continuum emission that is 
produced by thermal emission from disk heated by photospheric emission from the protostar and accretion (left panel of Fig.~\ref{Fig:L1527}). Early submm/mm interferometers had limited spectral line sensitivity, thus they primarily were able to detect continuum emission toward protostars (e.g., \ct{ChandlerSargent1993}). Moreover, the continuum sensitivity of interferometers has increased more quickly than spectral line sensitivity, making it the principle probe of protostellar disk studies prior to ALMA coming into full operations (\ct{Looney+2000,Harvey+2003,Jorgensen+2009,Maury+2010}). Even in the ALMA era, other instruments are still able to carry-out effective surveys due to their continuum sensitivity enabled by wider bandwidths to examine protostellar disks for increasingly large samples using smaller facilities (\ct{Tobin+2015a,Tobin+2015b,Segura-Cox+2018,Andersen+2019,Maury+2019}). But ALMA is also now providing increasingly large samples of protostellar systems in their continuum emission at sub-arcsecond wavelengths (\ct{Hsieh+2019a,Williams+2019}).
These studies are enabling both the continuum disk radii and disk masses to be characterized toward large samples that can be compared to the samples of protoplanetary disks (e.g., \ct{Ansdell+2016}).

Protostellar disks can also be detected in molecular line emission, tracing the gaseous component of the envelopes and disks. Molecular line emission then enables the kinematic structure of the inner envelope and disk to be examined, enabling differentiation between infall and rotation on these scales (e.g., \ct{Ohashi+1997,Yen+2013}). The most abundant molecule in the interstellar medium that can be readily detected in the submm/mm is carbon monoxide (CO). However, CO is so abundant that its emission lines are optically thick toward the dense cores harboring protostars, requiring that rarer species such as $^{13}$CO and C$^{18}$O be observed to overcome the opacity of the $^{12}$CO. These much less abundant species had been difficult to observe with sufficient angular resolution to detect and resolve protostellar disks prior to ALMA, aside from a few nearby systems (\ct{Jorgensen+2009,Takakuwa+2012,Tobin+2012,Yen+2013,Harsono+2014}). Furthermore, ALMA has taken the study of molecular line emission from protostellar disks using CO isotopologues to new heights (e.g., \ct{Yen+2014,Ohashi+2014,Aso+2015}), but at the same time species other than CO can now be readily detected in the envelopes and disks toward protostars (\ct{Sakai+2014b}), leading to studies of how chemical abundances may change as material is incorporated into the disk from the envelope (\ct{Sakai+2014a}).

For both continuum and molecular line emission, separating the envelope from the disk is challenging because the emission is entangled, overlapping in both projected position and velocity. To separate the envelope from the disk in the continuum, analysis of the data in the \textit{uv}-plane (the native units of interferometers) is required because the envelope and disk components can often be separated based on their amplitudes at different uv-distances. Separating the molecular line emission from the envelope and disk can be more difficult due to the disk and envelope having similar rotation profiles on the scale of the inner few hundred au. However, mapping the rotation curve of the emission can enable the infalling envelope portion to be separated from the disk (e.g., \ct{Yen+2013}) because the infalling envelope (with conserved angular momentum) and the disk should have different velocity profiles as a function of radius. The infalling envelope should have a v$_{\phi}$~$\propto$~R$^{-1}$, while the Keplerian disk should have v$_{\phi}$~$\propto$~R$^{-1/2}$, but even if the two slopes are distinct and have a clear `break radius', the disk may extend to larger radii than this break radius, but the envelope simply dominates the emission at that radius. Note that molecular line opacity from the infalling/rotating envelope could obscure emission from the disk depending on the viewing angle and velocity profiles, possibly making the break radius artificially low.

\subsubsection{Protostellar disk sizes from recent observations}

From the early studies, it became clear that resolving the disk from the surrounding protostellar envelope requires sub-arcsecond observations, which was challenging for early interferometers. A first high-resolution (sub-arcsecond) pilot survey was conducted by \citet{Maury+2010} with the IRAM/PdB interferometer at 1 mm, toward 5 Class 0 protostars. Surprisingly it suggested, although with a limited statistical significance, that the young protostellar disks remained mostly unresolved at scales $\sim 100$~au where hydrodynamical models were predicting these disks to extend (\ct{Bate2009}). Besides, CARMA was also used to survey nine protostellar systems at 100~au resolution in the Perseus molecular cloud, resolving several candidate disks in the dust continuum (\ct{Enoch+2011}). 

As a follow-up of the pilot study by \citet{Maury+2010}, the IRAM/PdBI interferometer was used to carry the Continuum and Lines in Young Protostellar Objects (CALYPSO) survey, providing sub-arcsecond observations of 16 nearby Class 0 protostars. They show that 11 of the 16 Class 0 protostars are better reproduced by models including a disk-like dust continuum component contributing to the flux at small scales, but less than 25\% of these candidate protostellar disks are resolved at radii $>$60 au. Including all available literature constraints on Class 0 disks at subarcsecond scales, they show that the CALYPSO statistics seem representative of Class 0 disks: most ($>$72\% in a sample of 26 protostars) Class 0 protostellar disks are small and detectable only at radii $<$60~au.

However, all disks detected in the continuum remain candidate disks because a key property of disks is to be rotationally supported. Prior to ALMA, the kinematics of only a few protostellar systems had been well-characterized, 
\citet{Tobin+2012} discovered a large $\sim 150$ au in radius, rotationally supported disk in the L1527 Class~0 protostar, using the gas kinematics traced by the $^{13}$CO emission obtained with CARMA. 
These were the first observations of a resolved, large, rotationally-supported Class 0 disk: they were suggested to be inconsistent with some disk formation models that consider strong magnetic braking leading to small young disks. 

\begin{figure}[ht]
\includegraphics[width=0.5\textwidth]{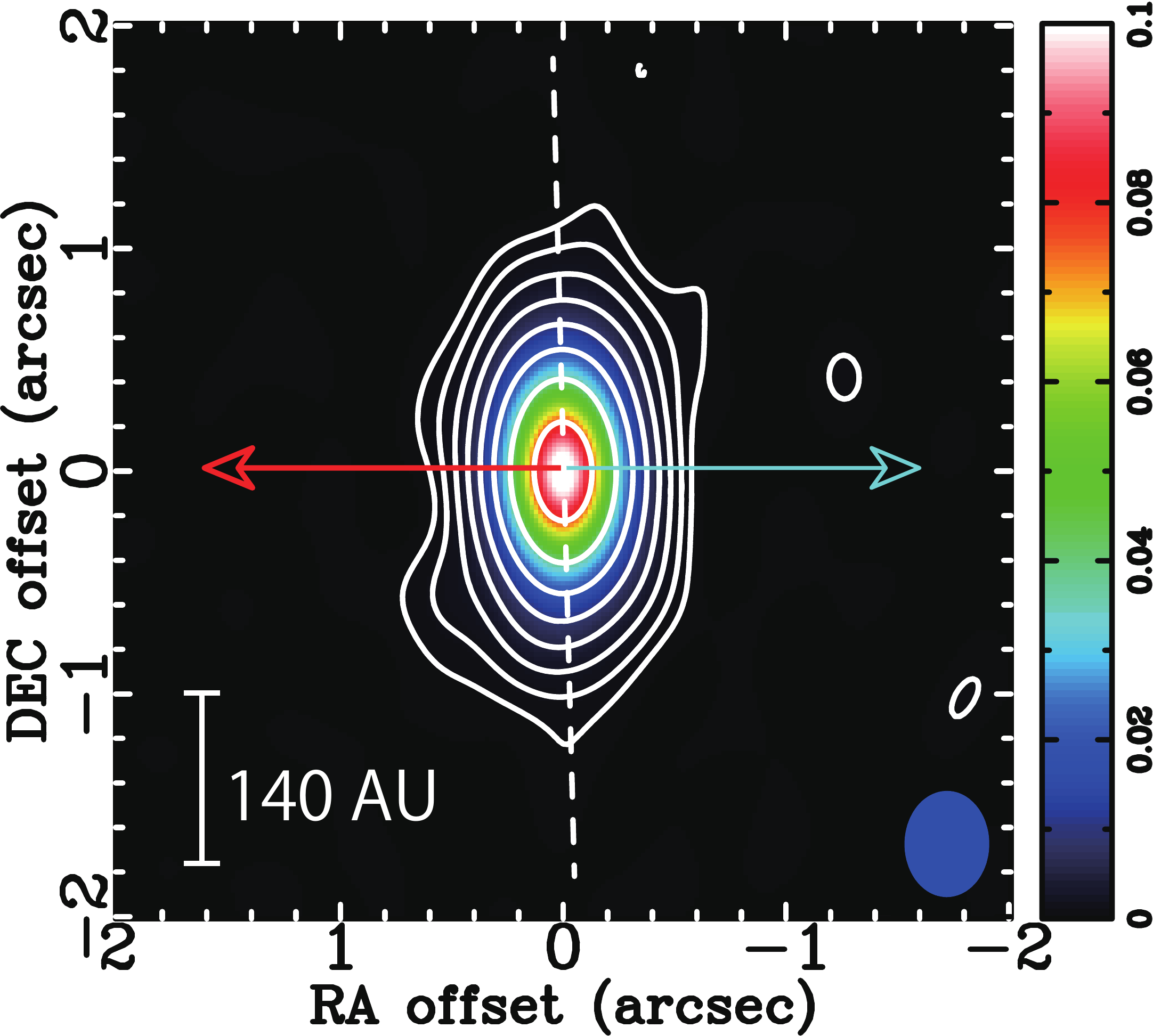}
\includegraphics[width=0.5\textwidth]{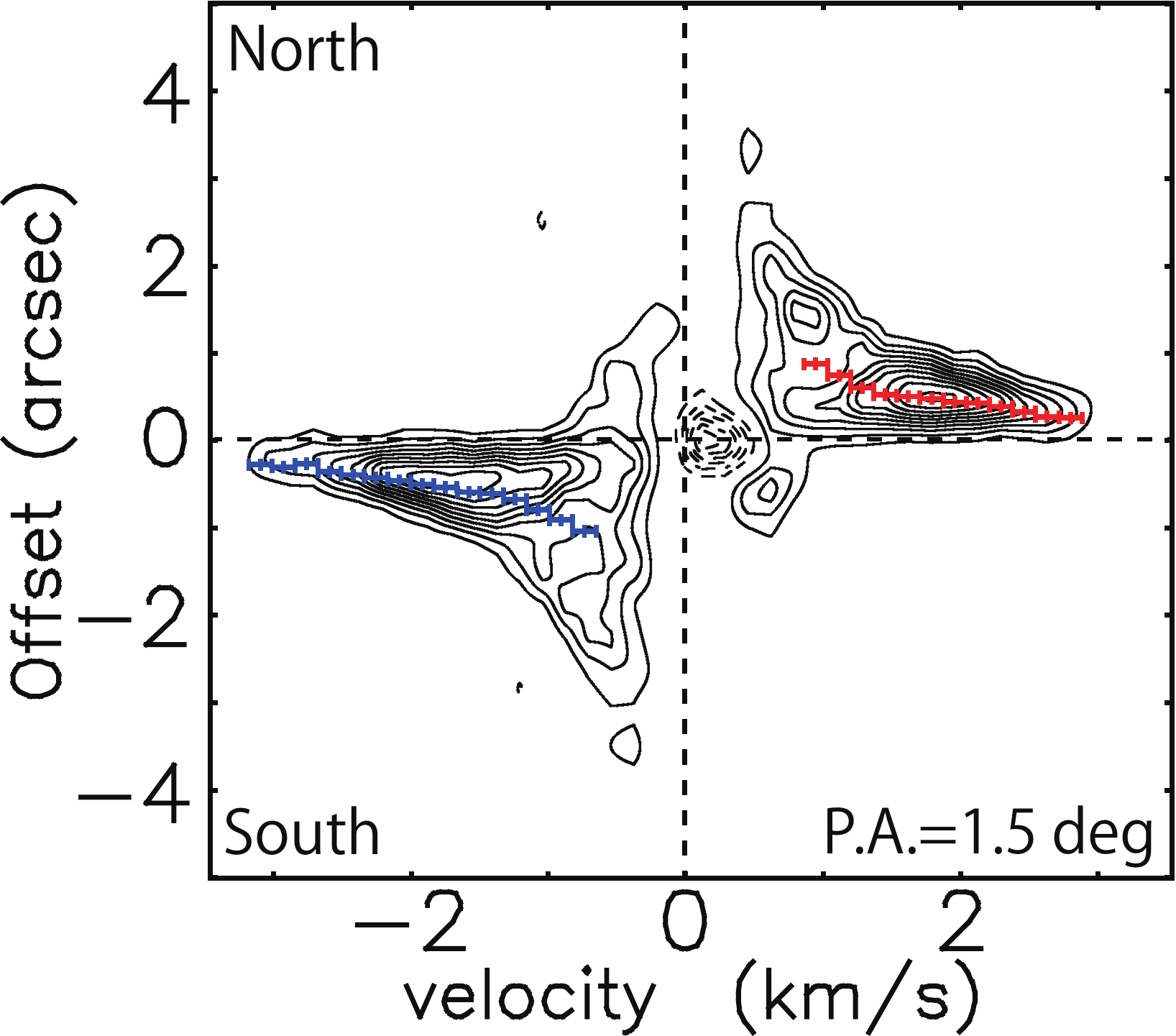}
\centerline{\includegraphics[width=0.5\textwidth]{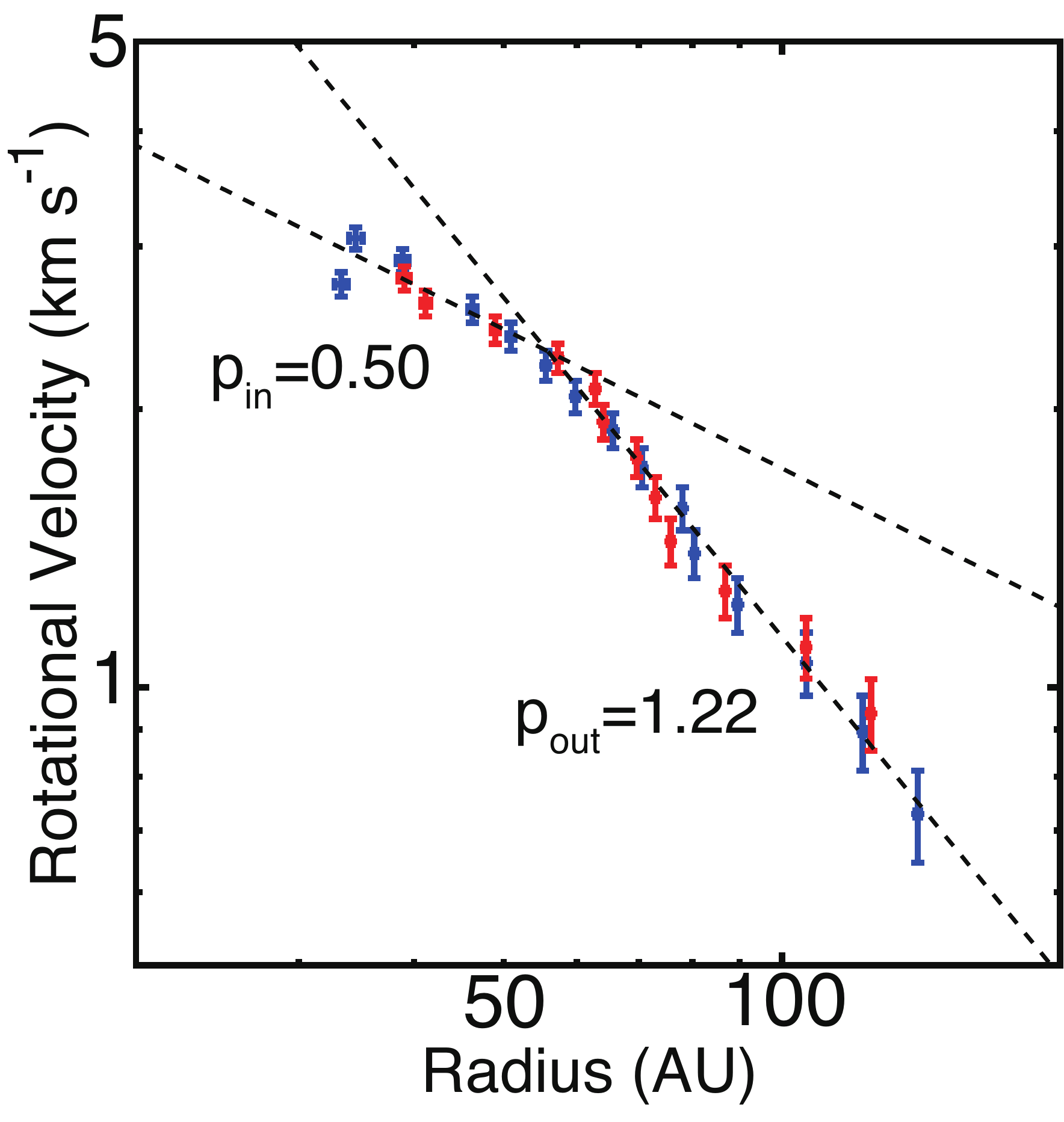}}
\caption{ALMA Observations and analysis of continuum and molecular line emission toward the Class 0 protostar L1527 IRS from \citet{Aso+2017}. The left panel shows the 1.3~mm dust continuum image resolving the dusty protostellar disk, the blue and red arrows denote the outflow direction from the protostar. The right panel shows a position-velocity diagram of the C$^{18}$O emission from the envelope and disk, with the spatial axis being extracted in the north-south direction along the major axis of the disk (left panel). The blue and red points represent fits to the centroid as a function of position and velocity in the PV diagram and are used to examine  the rotation profile of the emission. The bottom panel shows the fits to the position-velocity diagram in log-log space, highlighting the apparent change in the rotation curve from$v_{\phi}$~$\propto$~R$^{-1}$ to $v_{\phi}$~$\propto$~R$^{-0.5}$. Figures from \citet{Aso+2017}.}
\label{Fig:L1527}
\end{figure}

ALMA is now enabling the Keplerian disk radius to be detected for a number of systems (\ct{Murillo+2013,Ohashi+2014,Yen+2014,Aso+2015}), where the velocity profile clearly transitioned from $v_{\phi}$~$\propto$~R$^{-1}$ to $v_{\phi}$~$\propto$~R$^{-0.5}$, see Fig.~\ref{Fig:L1527}. With the ALMA C$^{18}$O observations (\ct{Ohashi+2014,Aso+2017}), the L1527 disk radius has been revised to $\sim56$ au. \citet{Maret+2020} find a Keplerian disk radius $\sim 90$au for this source using yet another molecular line (SO). In addition to the large radii measurements, a number of Class 0 protostars also seem to have very small disks with B335 appearing to have a disk radius as small as $\sim$3~au (\ct{Bjerkeli+2019}). 

However, the analysis of identifying the power law transition from the velocity profile requires several assumptions: that the molecular line emission 
is compact and that a velocity directly maps to a specific radius. Moreover, the break radius is also 
the point where the intensity of molecular line emission from the disk dominates from that of
the envelope and in the case of significant line opacity, the envelope can still effectively shield the emission from the disk. It has been found that even rarefied isotopologues like C$^{18}$O can be optically thick in protostellar envelopes and disks (\ct{VantHoff+2018}). Thus, observations of even more rarefied isotopolgues, like C$^{17}$O, are necessary if the tracer must be as optically thin as possible. 
Another disadvantage of rotation curve fitting to determine the Keplerian disk radius is that this method is using the highest velocity line emission which also tends to have the lowest signal-to-noise. This is because the lower velocities are significantly blended with envelope emission and are not reliably tracing the disk alone. 


As ALMA's capabilities have increased, the number of surveys being conducted toward protostellar systems has increased. This enables the distribution of disk radii to be characterized as an ensemble with significantly less bias than was present in the early ALMA observations that were carried out toward particular sources. The VLA/ALMA Nascent Disk and Multiplicity (VANDAM) Survey observed 93 protostars with the VLA at 9~mm in the Perseus molecular clouds, resolving candidate disks toward 18 systems (\ct{Tobin+2016a,Segura-Cox+2016,Segura-Cox+2018}). Statistical characterizations of disk radii for large samples are crucial for making further progress given that RSDs are confirmed in many Class 0 systems and a broad census is now required. The typical radii of disks have implications for the role that magnetic fields played in the collapse process. The protostars with very small disks either had to form in the presence of very little angular momentum (or have only recently collapsed), or magnetic fields removed a significant amount of angular momentum, preventing the formation of a large disk (\ct{Yen+2015,Maury+2018}).


\subsubsection{Protostellar disk masses from recent observations}

\begin{figure}[ht]
\centerline{\includegraphics[width=0.8\textwidth]{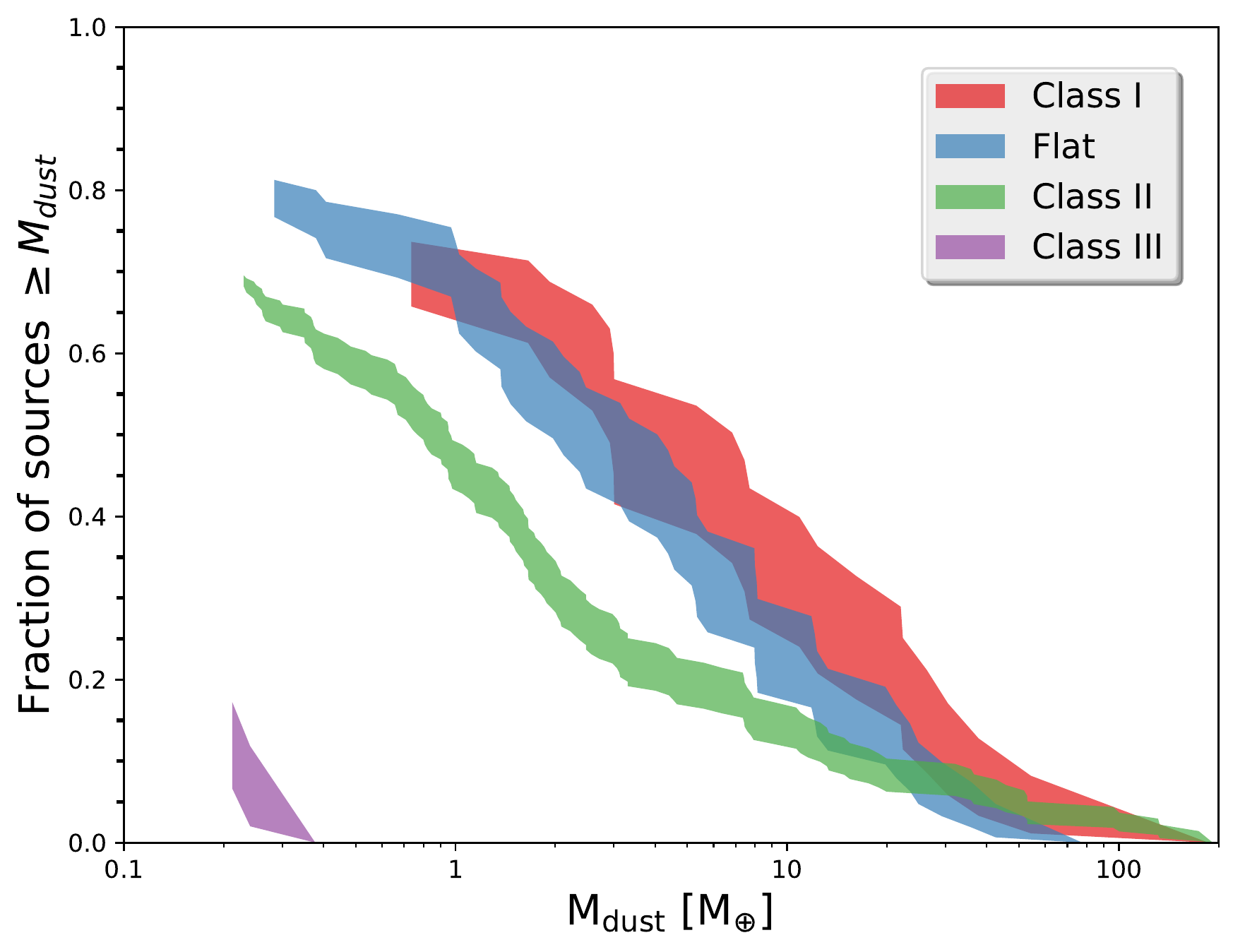}}
\caption{The cumulative distribution of dust mass for Ophiuchus disks around protostars of different infrared evolutionary states. The shading illustrates the 1$\sigma$ uncertainty at each mass and the colors indicate protostellar class. Figure from \citet{Williams+2019}.}
\label{Fig:odisea}
\end{figure}

One of the first large surveys of protostellar disks was toward the Ophiuchus region by \citet{Williams+2019}. This was part of the Ophiuchus DIsk Survey Employing ALMA (ODISEA) program which observed much of the population of young stars in Ophiuchus from Class I to Class III. The total sample size was 279, with 28 being Class I protostars and 50 being Flat Spectrum protostars (between Class I and Class II). They found that the logarithmic mean Class I dust disk mass to be 3.8 Earth masses, about 5 times larger than the mean Class II disk mass in Ophiuchus (Fig.~\ref{Fig:odisea}). The mean disk mass of Flat Spectrum protostars was in between the Class I and Class II, indicating a decrease of the typical disk mass with age in the Ophiuchus region. Assuming a typical dust to gas mass ratio of 1:100, the mean Class I disk gas mass would be 0.001~\msun\ which is quite low-mass when compared with the small disk survey of Perseus that found a mean Class 0 disk mass of 0.14~\msun (\ct{Tobin+2015b}).

Furthermore, in the VANDAM survey of protostars in Perseus with the VLA at 8~mm, \citet{Tychoniec+2018} find a typical disk mass of 0.14~\msun\ for Class 0 protostars and 0.05~\msun\ for Class I protostars. Similar measurements were found using a different analysis technique with lower-resolution data by \citet{Andersen+2019} and data at 1.3~mm, indicating that the measurements from Ophiuchus are very low compared to the Perseus region. Moreover, modeling of the Class I protostars in Taurus using CARMA data also finds larger Class I disk mass as compared to Ophiuchus (\ct{Sheehan+2017b}). Thus, we can conclude that disk masses toward protostars are systematically larger than the more-evolved Class II disks, but it is unclear if the mean protostellar disk masses are similar between regions or if we should expect them to be similar, given the differences seen in the Class II populations with similar ages (\ct{Williams+2019,Cazzoletti+2019}). Such large differences between regions may point to the initial conditions that lead to the formation of a particular star forming region playing a role in setting the ensemble properties of protostellar disks. Observations tend to suggest the youngest disks are both more compact and more massive than their evolved counterparts, which may facilitate the early planet formation, at scales yet unprobed by most current observations.



\subsubsection{Protostellar disks chemistry}


While a ubiquitous tracer of protostellar disks as compared to the envelope remains elusive, emission from different molecules can distinctly trace the envelope and disk in some systems. For example, \citet{Sakai+2014a} found that the envelope of L1527 IRS emitted in CCH and c-C$_3$H$_2$, while the disk was devoid of emission in these molecules. The disk instead appears in emission of the SO molecule, which in turn was not being emitted by the envelope. Then, several tracers appeared to be present in both the disk and envelope, like CS, H$_2$CO, and C$^{18}$O. The authors argued that there was a clear change in the chemical abundances of certain molecules as the envelope material became incorporated into the disk. The drop in abundance of molecules like CCH and c-C$_3$H$_2$ and increase in SO was argued to result from a shock generated by the infall of material onto the disk (\ct{Yen+2014}) or a centrifugal barrier (\ct{Sakai+2014b}). However, the clear distinction of the different molecular species between the disk and envelope has not yet been confirmed for a large number of sources, only L1527 IRS and TMC1A, with TMC1A being a less clear case.




\subsection{Theoretical \& numerical development on disk formation}
\label{S.TheoryLowMass}

Driven by these striking discoveries in high-resolution observations of young disks, plenty of progress are also made on the theoretical and numerical side, as to explain how such disks with the observed properties are formed. In this section, we highlight the impact of non-ideal MHD effects and perturbations of the parent core on the formation and evolution of disks. 

The main physical processes involved in protostellar collapse are MHD, self-gravity, radiation transfer, and non-ideal MHD effects. The governing equation can be described as follow,
\begin{equation}
{\partial \rho \over \partial t} + \nabla \cdot (\rho \mathbf{v}) = 0~,
\end{equation}
\begin{equation}
{\partial \mathbf{v} \over \partial t} + (\mathbf{v}\cdot\nabla)\mathbf{v} = -{\nabla P \over \rho} + {1 \over 4 \pi \rho} (\nabla \times \mathbf{B}) \times \mathbf{B} - \nabla {\rm \Phi}_g~,
\end{equation}
\begin{equation}\label{Eq:inductB}
{\partial \mathbf{B} \over \partial t} = \nabla \times (\mathbf{v \times B}) - \nabla \times \left\{\eta_{\rm O}\nabla \times \mathbf{B} + \eta_{\rm H}(\nabla \times \mathbf{B}) \times {\mathbf{B} \over B} + \eta_{\rm AD}{\mathbf{B} \over B} \times \left[(\nabla \times \mathbf{B}) \times {\mathbf{B} \over B} \right]\right\}~,
\end{equation}
\begin{equation}
\nabla^2 {\rm \Phi}_g = 4\pi G \rho~,
\end{equation}
\begin{equation}\label{Eq:divB}
\nabla \cdot \mathbf{B} = 0
\end{equation}
where $\rho$, $P$, $\mathbf{v}$, $\mathbf{B}$, and ${\rm \Phi}_g$ are gas density, gas pressure, gas velocity, magnetic field, and gravitational potential, respectively, and $\eta_{\rm O}$, $\eta_{\rm H}$, and $\eta_{\rm AD}$ are the Ohmic, Hall and ambipolar diffusivities, respectively. The set of equations are closed by the energy equation, which can be approximated by a barotropic equation of state (EOS; e.g., \ct{Machida+2008}), or directly solved by full radiation transfer (e.g., \ct{Tomida+2013}). Both Eulerian grid-based code (e.g., \texttt{RAMSES}; \ct{Masson+2016}) and Lagrangian SPH code (e.g., \texttt{PHANTOM}; \ct{Wurster+2016}) have been developed to solve the above equations numerically, as to help understand various aspects of the formation and evolution of young disks.

Note that the divergence free condition of the magnetic field (Eq.~\ref{Eq:divB}) does not necessarily hold in numerical simulations, and has to be maintained through either divergence cleaning (\ct{Dedner+2002}) or the more self-consistent method of constrained transport (\ct{EvansHawley1988}).

\subsubsection{Ideal MHD limit \& magnetic braking catastrophe}
\label{S.ideal}

Early studies of disk formation in collapsing cores usually assume that magnetic fields are well coupled to the bulk neutral fluid. In such an ideal MHD limit, the induction equation Eq.~\ref{Eq:inductB} can be simplified as,
\begin{equation}
{\partial \mathbf{B} \over \partial t} = \nabla \times (\mathbf{v \times B})~.
\end{equation}
Therefore, as the gravitational collapse continues, the magnetic field is dragged inward by the neutral gas to form a split monopole configuration, with envelope material sliding freely along the highly pinched magnetic field lines, flattening the equatorial region into a ``pseudo-disk" structure (\ct{GalliShu1993a,GalliShu1993b}). The pinching of magnetic field lines across the pseudo-disk results in a strong magnetic torque ($\propto \varpi B_z {\partial B_\phi \over \partial z}$, where $\varpi$ is the cylindrical radius, and $B_z$ and $B_\phi$ are the poloidal and toroidal magnetic field components, respectively) that transports most of the angular momentum away from the pseudo-disk plane and brakes the gas rotation. Due to a lack of rotational support, gas continues to collapse to the central stellar object along the pseudo-disk, without forming sizable disks that can be compared with observations. This is the so-called magnetic braking ``catastrophe" in the ideal MHD limit that has been discovered early on by different groups (\ct{Tomisaka2000,Allen+2003,MellonLi2008,HennebelleFromang2008}). \citet{Allen+2003} showed that even for $\lambda$ as high as 10, 
no disk would form. While this sounds at first surprising, it is worth stressing that this corresponds to the analytical expectation as stated by Eq.~\ref{eq_br3}. More quantitatively, \citet{MellonLi2008} performing axisymmetric 2D collapse, reported that for $\lambda$ as high as 10, no disk forms while the disk gets affected by magnetic field for values at high as 100. Using 3D adaptive mesh refinement simulations, \citet{HennebelleFromang2008,HennebelleTeyssier2008} obtained similar though slightly smaller values, i.e. a threshold for disk formation for $\lambda$ around 5 and disk influence around $\lambda \simeq 50$. Very similar values are also reported by \citet{PriceBate2007}.

Later, 3D simulations adopting sink particle treatment (\ct{Zhao+2011,Seifried+2011,Cunningham+2012}) discovered that, in addition to the magnetic braking, magnetic pressure dominated structures developed in the stellar vicinity by magnetic interchange instability (\ct{Parker1979,StehleSpruit2001}) further block the accretion flow from rotating freely around the central object. The formation of such structures is a result of the decoupled magnetic flux left behind in the cell by the accreted matter onto the sink particle. Despite being somewhat artificial, such a treatment of the magnetic flux around the sink particle is in fact a crude representation of the matter-field decoupling expected at high densities ($\gtrsim10^{12}$~cm$^{-3}$; \ct{Nakano+2002,KunzMouschovias2010}). The magnetic interchange instability driven by magnetic flux redistribution is later confirmed by the non-ideal MHD study including ambipolar diffusion and Ohmic dissipation (\ct{Krasnopolsky+2012}). 
In contrast, a few SPH codes (e.g., \ct{Wurster+2014}) directly delete the magnetic flux associated with the accreted matter; as a result, such a phenomena is not present in the corresponding simulations, and the formation of RSDs becomes possible even in the ideal MHD limit (e.g., \ct{Wurster+2017}). 

Note that the magnetic braking ``catastrophe" is the most severe in the axisymmetric case when the magnetic field and rotation axis are aligned; non-axisymmetric perturbations such as magnetic misalignment and/or turbulence can alleviate the problem to certain extent, which will be discussed in \S~\ref{S.IC_impact}.

\subsubsection{Non-ideal MHD effects}
\label{S.non-ideal}

In reality, dense cores are only slightly ionized (with typical electron fractional abundance of $\sim$10$^{-7}$; \ct{BerginTafalla2007}), hence the flux-freezing condition in the ideal MHD limit is not strictly satisfied, and decoupling of magnetic fields from the bulk neutral fluid should commonly occur in the collapsing core. Therefore, it is natural and necessary to include non-ideal MHD effects (AD, Hall effect, and Ohmic dissipation; Eq.~\ref{Eq:inductB}) in the study of the formation and evolution of young disks. Generally, AD can be efficient in both the inner and outer envelopes; Hall effect mainly dominates the inner envelope ($\sim$100~au scale); and Ohmic dissipation only becomes important in the high density disk itself, provided that disk is already formed. We summarize the recent progress of non-ideal MHD studies below.

\vspace{0.5cm}
\noindent $\bullet$ {\it Ohmic dissipation}
\vspace{0.5cm}

\noindent Many early studies of disk formation focus on the Ohmic dissipation, which operates as a diffusive term in the induction equation to weaken the electromotive force (EMF) by dissipating the total electric current. \citet{Shu+2006} adopted semi-analytically models to investigate the role of a constant Ohmic resistivity in collapsing magnetized cores. They demonstrated that Ohmic dissipation renders magnetic field lines to become asymptotically straight and uniform close to the central stellar object, and suggested that disk formation may be enabled by an enhanced resistivity of the infalling gas by at least 1 order of magnitude larger than the classic microscopic value. Later, \citet{Krasnopolsky+2010} using numerical simulations, confirms that an enhanced resistivity of the order 10$^{19}$~cm$^{2}$~s$^{-1}$ (can vary by a factor of a few for different core magnetization) is necessary to enable the formation of RSDs of 100~au. 

\citet{DappBasu2010} adopted thin-disk approximation with Ohmic dissipation to simulate the protostellar collapse, and showed that magnetic braking is ineffective within the first core and a tiny disk of about 10 stellar radii is able to form shortly after the second collapse. They speculated that such a disk may grow in radius into the Class II phase if both AD and Ohmic dissipation are included. \citet{MachidaMatsumoto2011} confirmed that a compact but massive Keplerian disk can form from the first core with Ohmic dissipation alone. Later, \citet{Machida+2011} evolves the system longer in time with the help of a central sink, and claims that the initial small Keplerian disk can grow to $\gtrsim$100~au after most of the infalling envelope has been accreted by the end of the main accretion phase. However, both work (\ct{MachidaMatsumoto2011,Machida+2011}) simplified the Ohmic dissipation term in the induction equation assuming a constant resistivity, yet a varying resistivity from \citet{Machida+2007} is adopted. Therefore, the reduction of the EMF through Ohmic dissipation of the electric current is overestimated by neglecting the contribution from a vertical gradient of Ohmic resistivity ($\nabla \eta_{\rm O}$) across the pseudo-disk. Indeed, \citet{Tomida+2013} corrected such a treatment and conducted resistive MHD simulations with full radiation transfer, the system evolves to about one year after the formation of the second (protostellar) core, when a small circumstelalr disk ($\lesssim$0.35~au) is formed due to Ohmic dissipation, similar to the result of \citet{DappBasu2010}.

\vspace{0.5cm}
\noindent $\bullet$ {\it Ambipolar diffusion}
\vspace{0.5cm}

\noindent Ambipolar diffusion is also a diffusive process that allows the bulk neutral fluid to drift relative to the charged species that are relatively well coupled to the magnetic field. Because of the drag force of the neutrals exerted onto the charged species, the net effect of the ambipolar drift is to relax the tendency of bending of magnetic field lines.
The ambipolar drift velocity $\mathbf{v}_{\rm d}$ can be expressed as,
\begin{equation}
\mathbf{v}_{\rm d} \equiv \mathbf{v}_i - \mathbf{v}_n = \eta_{\rm AD} {(\nabla \times \mathbf{B}) \times \mathbf{B} \over B^2}~,
\end{equation}
in which $\mathbf{v}_{\rm i}$ and $\mathbf{v}_n$ are the velocities of the dominant charged species (e.g., ions or charged grains) and the neutrals, respectively.

Early work incorporating AD in the study of protostellar collapse and disk formation use semi-analytical solutions (\ct{Contopoulos+1998,KrasnopolskyKonigl2002}) of self-similar collapse models or 1D simulation with thin-disk approximation (\ct{CiolekKonigl1998}). They demonstrated that, after the point mass formation, AD efficiently decouples magnetic field from the infalling neutral matter in the inner envelope, causing the infall motion of the field lines to stall abruptly within a characteristic radius. Such a decoupling process drives a hydrodynamic C-shock that moves radially outward into the infalling flow, enhancing the magnetic field strength in the inner envelope; this so-called ``AD-shock" is first demonstrated in \citet{LiMcKee1996}. \citet{KrasnopolskyKonigl2002} found that magnetic braking is actually enhanced behind the AD-shock, which further suppresses disk formation. This result is later confirmed by the 2D simulation of \citet{MellonLi2009}, who adopted a simple parametrized ambipolar diffusivity with relatively large ion density coefficient (\ct{Shu1991}) and suggested that disk formation only becomes possible in dense cores with weak magnetic field and/or low cosmic ray ionization rate. \citet{Li+2011} extended the study by using magnetic diffusivities computed from a minimal chemical network including charged dust grains (\ct{Nishi+1991}), and confirmed the enhancement of magnetic braking in the post AD-shock region and the suppression of disk formation. Note that their choice of grain size distributions results in an ambipolar diffusivity not far from the simple parametrized values of \citet{MellonLi2009}. \citet{Dapp+2012} combines AD and Ohmic dissipation into an effective resistivity in their thin-disk framework and explored different grain sizes for the resistivity computation. They corroborated their previous conclusion of \citet{DappBasu2010} that only very small disks of $\sim$10 stellar radii are able to form around the second core. However, the small disk radius is likely caused by their somewhat high cosmic-ray ionization rate (5$\times$~10$^{-17}$~s$^{-1}$) and slow rotation speed of the initial core.

More recently, the role of AD in disk formation has been revisited, particularly through 3D simulations with scrutiny of the ionization chemistry and magnetic diffusivities. Most grid-based simulations show that AD greatly promotes disk formation while the SPH simulations do not seem to suggest so. 

\citet{Masson+2016} showed that AD alone can result in RSDs of $\sim$20--30~au radius in strongly magnetized cores ($\lambda$$\sim$2) and $\sim$40--80~au radius from moderately magnetized cores ($\lambda$$\sim$5). They found that the magnetic field strength is saturated at a level of 0.1~G within a magnetic diffusion barrier, which reduces the magnetic braking efficiency in the inner envelope. Basically, the ambipolar diffusion becomes efficient at high density and operates on a time scale shorter than the free-fall time.
Such a magnetic diffusion barrier is conceptually similar to the AD-shock (\ct{Li+2011}; see also \ct{Lam+2019}), but is less obstructive to the infalling flow and not obviously enhancing the magnetic field strength across the barrier. The discrepancy likely originates, as pointed out in \citet{Zhao+2016}, from the grain size distribution in their chemical network (\ct{Marchand+2016}), which enhances the ambipolar diffusivity by a factor of $\sim$10 as compared to that used in \citet{Li+2011} and \citet{MellonLi2009}. \citet{Hennebelle+2016} extended this work and derived semi-analytically an expression for the early disk radius, based on the magnetic self-regulation assumption that inward flux advection by infall is balanced by the outward flux diffusion by AD. They concluded that the typical radius of Class 0 disks should be around $\sim$20~au, with weak dependence on the initial magnetic field strength and core rotation (see also \S~\ref{S.DiskSize}). 

\citet{Tomida+2015} extended their previous radiation MHD study (\ct{Tomida+2013}) by including AD. They confirmed that while Ohmic dissipation only becomes important in the highest densities and results in very small disks, ambipolar diffusion of magnetic fields can operate at lower densities and enable sizable disks to form. Note that their ambipolar diffusivity (by choosing single-sized 0.1~$\mu$m grain) is even a few times higher than that of \citet{Masson+2016}; however, the resulting disk radius is as small as $\sim$5~au, possibly because of the slow initial core rotation with angular speed of $2.4\times10^{-14}$~s$^{-1}$ that corresponds to a low $\beta_{\rm rot}\approx3.3\times10^{-3}$ and a relatively early time when the simulation is stopped.

\begin{figure}[ht]
\centerline{\includegraphics[width=1.0\textwidth]{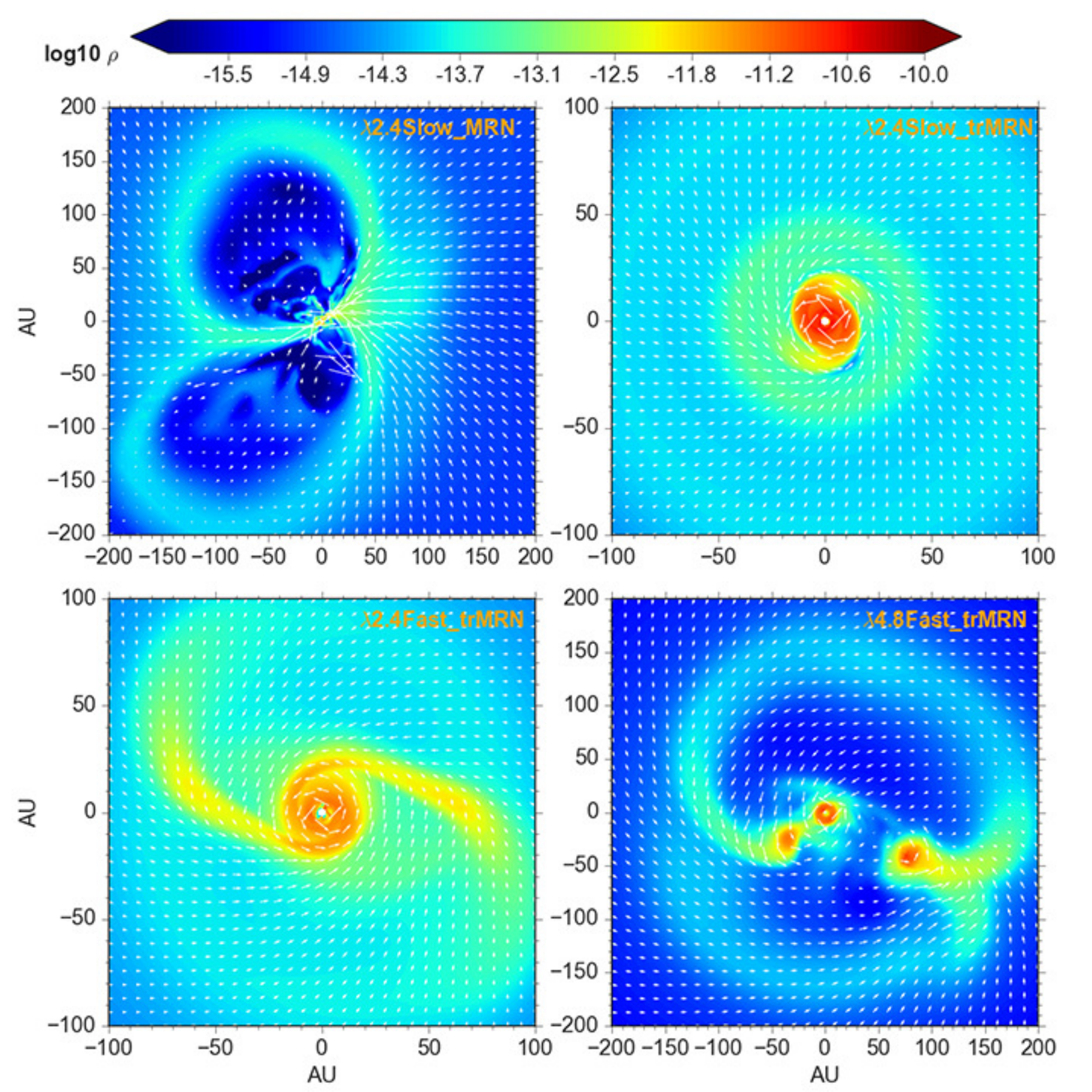}}
\caption{Distribution of mass density for different non-ideal MHD models (AD). Top left: VSGs present, strong B field (mass-to-flux ratio $\lambda$=2.4), and slow rotation ($\beta_{\rm rot}$=0.025) model. No rotationally supported disks, only magnetically dominated structure $-$ DEMS (decoupling-enabled magnetic structures). Top right: VSGs absent, strong B ($\lambda$=2.4), and slow rotation ($\beta_{\rm rot}$=0.025) model. Rotationally supported disk. Bottom left: VSGs absent, strong B ($\lambda$=2.4), and fast rotation ($\beta_{\rm rot}$=0.1) model. Large spiral structure. Bottom right: VSGs absent, weak B ($\lambda$=4.8), and fast rotation ($\beta_{\rm rot}$=0.1). Triple stellar system.}
\label{Fig:ADDisk}
\end{figure}

\begin{figure}[ht]
\centerline{\includegraphics[width=1.0\textwidth]{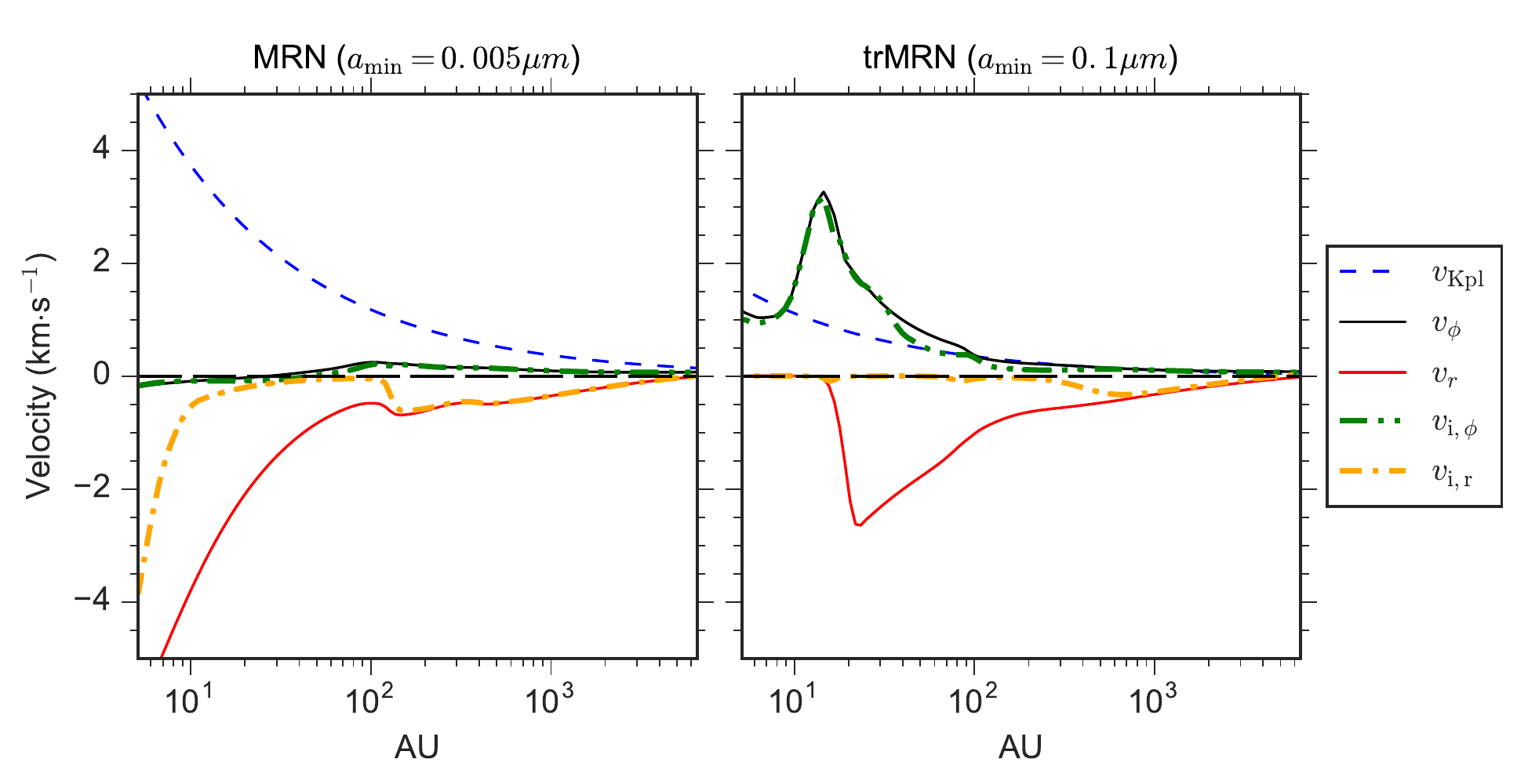}}
\caption{Infall and rotational velocities for the neutrals and ions along the pseudo-disk for models with MRN grains (left panel) and trMRN grains (right panel). The ions in the MRN model only start to decouple from the neutrals within the inner $\lesssim$100~au, while the ion velocity in the trMRN model is almost vanishing within $\lesssim$1000~au. The initial mass-to-flux ratio is $\sim$2.4 for both models.}
\label{Fig:driftVel}
\end{figure}

\citet{Zhao+2016} investigated the effect of grain size distribution and cosmic-ray ionization rate on the magnetic diffusivity and disk formation. They found that the population of very small grains (VSGs: $\sim$1~nm to few 10~nm) in fact dominate the fluid conductivity, i.e., the coupling of the bulk neutral matter to the magnetic field, in the collapsing core. Removing such a VSG population would enhance the ambipolar diffusivity by $\sim$1--2 orders of magnitude and reduce the amount of magnetic flux dragged in by the collapse even at the first core phase. As a result, tens-of-au RSDs are able to form. They also claimed that cosmic-ray ionization rate at the core-scale should be below 2--3~$\times$~10$^{-17}$~s$^{-1}$ in order for AD to be efficient in enabling disk formation. \citet{Zhao+2018a} extended this 2D work into 3D and found that (see Fig.~\ref{Fig:ADDisk}), by setting a minimum grain size $a_{\rm min}=0.1~\mu$m, the outward ambipolar drift speed of magnetic field lines is greatly enhanced that it almost cancels out the infall speed of the neutrals along the 10$^2$--10$^3$~au scale pseudo-disk (see also Fig.~\ref{Fig:driftVel}). They also suggested that lowering magnetic field strength and/or increasing core rotation can promote the formation of spiral structures as well as companion clumps near the centrifugal barrier where infalling materials from the envelope pile up. 

On the other hand, recent SPH studies seem to suggest that AD is not an efficient mechanism to enable disk formation. \citet{Tsukamoto+2015a} included AD and Ohmic dissipation and evolved the system up to the second collapse, when the RSD is as small as $\lesssim 1$~au. It is largely consistent with the result of \citet{Tomida+2015}. However, due to a lack of sink particle treatment, their simulation did not continue into the main accretion phase; hence the long-term effect of AD on disk evolution is unclear. In comparison, \citet{Wurster+2016} demonstrated that disk formation with AD alone is still suppressed as in the ideal MHD limit, although their initial conditions are similar to \citet{Masson+2016} and their ambipolar diffusivity is slightly larger (with 0.1~$\mu$m grain) than \citet{Masson+2016}. The discrepancy may originate from the numerical method of the ambipolar drift of magnetic fields by SPH particles.

It is worth noting that in the recent grid-based simulations, AD-shock is less obvious or mostly absent, which is likely a result of the enhanced ambipolar diffusivity used in these studies. Increasing the overall efficiency of AD causes the magnetic field to start decoupling from the collapsing flow gradually at the envelope scale and reduces the amount of magnetic flux arrived at the stellar vicinity. Hence the abrupt decoupling process responsible for the formation of the AD-shock is greatly suppressed; this phenomenon has been discussed in various previous studies (\ct{KrasnopolskyKonigl2002,Zhao+2018a,Lam+2019}).

\vspace{0.5cm}
\noindent$\bullet$ {\it Hall effect}
\vspace{0.5cm}

\noindent Hall effect has been brought into attention very recently, which, unlike AD and Ohmic dissipation, is a dispersive process that drifts magnetic field lines along a direction orthogonal to the bending direction of the magnetic field. The Hall drift velocity $\mathbf{v}_{\rm H}$ can be expressed as,
\begin{equation}
\mathbf{v}_{\rm H} = \mathbf{v}_e - \mathbf{v}_i = -\eta_{\rm H} {\nabla \times \mathbf{B} \over B}~,
\end{equation}
in which $\mathbf{v}_e$ and $\mathbf{v}_i$ are typically the velocities of the electrons and the ions, respectively, for a simple electron-ion plasma. For instance, Hall effect can drift the radially pinched magnetic field lines along the azimuthal direction, resulting in a magnetic tension force directed in the azimuthal direction that spins up/down the neutral fluid azimuthally.

\citet{Krasnopolsky+2011} was the first to numerically study the Hall effect in protostellar collapse and disk formation. They adopted constant values of Hall coefficient ${\eta_{\rm H} \over B}$ and showed that, depending on the orientation of the magnetic field with respect to the rotation axis ($\mathbf{\Omega \cdot B}$ $>0$ or $<0$), Hall effect can spin-up/down the inner part of the collapsing flow and produce both regularly rotating and counter-rotating RSDs. However, they suggested that the typical Hall coefficient in dense cores (e.g., using the standard MRN grain size distribution; Mathis-Rumpl-Nordsieck; \ct{Mathis+1977}) is about one order of magnitude smaller than that needed for disk formation by Hall effect. The result is later confirmed by \citet{Li+2011} with more realistic Hall diffisivity computed from the minimal chemical network.  \citet{BraidingWardle2012a,BraidingWardle2012b}) developed semi-analytical models and found that flipping the sign of $\mathbf{\Omega \cdot B}$ (or equivalently the sign of $\eta_{\rm H}$) can change the radius of RSDs by an order of magnitude. 

More recently, 3D SPH simulations further highlight the role of Hall effect in disk formation. \citet{Tsukamoto+2015b} included all three non-ideal MHD effects and concluded that Hall effect is the dominant mechanism which leads to a bimodal formation of protostellar disks, i.e., disk formation is promoted when $\mathbf{\Omega \cdot B}<0$ (anti-aligned) but suppressed when $\mathbf{\Omega \cdot B}>0$ (aligned). However, their simulation stopped shortly after the protostar formation, it remains unclear how such envelope-disk system evolves in the main accretion phase. \citet{Wurster+2016} evolves the systems somewhat longer in time ($\sim$5~kyr after the protostar formation) and arrives at a similar conclusion that disk formation is only possible in the anti-aligned configuration ($\mathbf{\Omega \cdot B}<0$). They also suggested that the disk radius is larger ($\sim$38~au) when including only Hall effect, and smaller ($\sim$13~au) when including all three non-ideal MHD effects, implying that AD hinders the efficiency of Hall spin-up/down. Note that such a bimodality of disk formation seems to contradict the result of \citet{Krasnopolsky+2011} that either polarity of magnetic fields can produce RSDs. 

\citet{Marchand+2018}, implemented Hall effect in \texttt{RAMSES} and raised the issue of violation of angular momentum conservation in collapse simulations including Hall effect. However, the problem likely comes from their numerical algorithm (\ct{Lesur+2014}), rather than an intrinsic property of Hall effect. In fact, the angular momentum conservation has been improved in their subsequent work (\ct{Marchand+2019}) by removing the whistler wave speed in the Riemann solver for non-magnetic variables.

Note that, similar to AD, Hall effect is also sensitive to chemistry and microphysics, especially the grain size distribution. The discrepancy between the results of recent studies and that of \citet{Li+2011} is likely due to different choices of grain sizes, as suggested by \citet{Zhao+2018b}. We discuss the effect of chemistry and microphysics on the magnetic diffusivities in the next section.

\vspace{0.5cm}
\noindent $\bullet$ {\it Link to Microphysics: Cosmic-ray Ionization Rate and Grain Size Distribution}
\vspace{0.5cm}

\noindent The efficiency of non-ideal MHD effects relies heavily on the value of magnetic diffusivities in the induction equation ($\eta_{\rm O}$, $\eta_{\rm AD}$, and $\eta_{\rm H}$ in Eq.~\ref{Eq:inductB}), which are determined by chemistry and microphysics (\ct{OppenheimerDalgarno1974,UmebayashiNakano1990,Nakano+2002}), especially the ionization rate and grain size distribution.

As most of the interstellar UV-radiation are attenuated at relatively high column densities (visual extinction Av$>$4~mag; \ct{McKee1989}), cosmic rays become the main source of ionization in dense cores. The typical level of cosmic-ray ionization rate $\zeta^{\rm H_2}$ in dense cores inferred from chemical analysis ranges from a few 10$^{-18}$~s$^{-1}$ to a few 10$^{-16}$~s$^{-1}$ (\ct{Caselli+1998,Padovani+2009}). Usually, $\zeta^{\rm H_2}\approx10^{-17}$~s$^{-1}$ is considered as the ``standard" cosmic-ray ionization rate in dense cores. Note that the ionization rate appears to be systematically higher in massive protostellar envelopes than in low mass ones (\ct{Padovani+2009}), possibly due to additional ionization from the young stellar object (\ct{Padovani+2015,Padovani+2016}). Non-ideal MHD studies including only AD and Ohmic dissipation show that a slightly higher than standard $\zeta^{\rm H_2}$ $\gtrsim$2--3$\times$10$^{-17}$~s$^{-1}$ at the core scale can already suppress the formation of RSDs of tens-of-au (\ct{Zhao+2016}), leaving only small RSDs of $\lesssim$10~au radii (\ct{Dapp+2012}). Further numerical explorations by including Hall effect are necessary to determine the constraint on $\zeta^{\rm H_2}$.

Furthermore, the collapsing dense core is not a simple electron-ion-neutral medium, charged grains play an important role in the fluid conductivity and magnetic diffusivity. Given a dust to gas mass ratio of $\sim$0.01, smaller grain sizes provide more surface area for electrons and ions to recombine, which tends to reduce the ionization fraction ($\propto a$, where $a$ is the grain radius). However, since smaller grains (especially VSGs) are relatively well-coupled to the magnetic field and at the same time have reasonably large collisional cross-sections with the neutral molecules, a large population of charged VSGs can dominate the fluid conductivity (instead of ions and electrons). Studies by different groups have demonstrated the pivotal role of VSGs in the magnetic diffusivities (\ct{Dzyurkevich+2017,Zhao+2018b,Koga+2019}). The discrepancies among existing numerical studies on magnetic braking and disk formation largely originates from the choices of the grain size distributions with or without VSGs (\ct{Li+2011,Masson+2016}).
Therefore, it is crucial to constrain the abundance of VSGs in collapsing dense cores. In fact, recent Galactic cold core surveys (\ct{Tibbs+2016}) searching for emission of spinning dust grains (\ct{DraineLazarian1998}) suggest a depletion of $\lesssim$10~nm grain in their sample of dense cores. Analytical and numerical models have also shown that the depletion of VSGs can be very rapid in cold dense environment (\ct{Ossenkopf1993,Hirashita2012}), where the VSGs are efficiently absorbed onto bigger grains (e.g., \ct{Kohler+2012}), a process in analogous to the molecular freeze-out (\ct{Caselli+1999}). 

On the other end of the grain size distribution, recent discovery of ``coreshine" (light scattered by large dust grain up to 1~$\mu$m; \ct{Pagani+2010,Steinacker+2010}) from nearby dense cores implies that the maximum grain size in dense cores can already reach micron sizes. Note that the maximum grain size has little effect on the conductivity and magnetic diffusivities, as long as the sub-micron grains are still present (\ct{Zhao+2016}). In summary, the grain size distribution in dense cores may already deviate from the standard MRN size distribution inferred from the diffuse interstellar medium; obtaining a more realistic grain size distribution for dense cores would have profound impact on the magnetic flux transport in protostellar envelope and on the formation of protostellar disks.

\subsubsection{Dependence of disk size on physical conditions}
\label{S.DiskSize}

One common result in the non-ideal MHD studies of disk formation is that the size of the central compact disk does not strongly depends on the initial magnetic field strength and cloud rotation (\ct{Hennebelle+2016}). However, the extent of the extended spiral structures is instead sensitive to such initial parameters of the molecular cloud (\ct{Zhao+2018a}). In general, weaker cloud magnetization and/or faster rotation promote the formation of spiral structures and multiple systems (\ct{Tomida+2017,Zhao+2018a,WursterBate2019}; see also \S~\ref{Chap.GIFrag}). 
In comparison to the disk morphology of non-magnetized collapse simulations (e.g., \ct{Matsumoto+2003}), the region of spiral structures in non-ideal MHD simulation are subject to persistent magnetic braking, therefore, such structures can be viewed as a degeneration of the outer disks in non-magnetized simulations. On the other hand, the relatively small size of the central compact disk is likely due to the non-ideal MHD induced disk stationarity (\ct{Hennebelle+2016}).


\vspace{0.5cm}
\noindent $\bullet$ {\it Magnetic diffusion and stationarity}
\vspace{0.5cm}

\noindent When non-ideal MHD is included, magnetic field is imperfectly coupled to the gas, the magnetic intensity at high density is not necessarily increasing with the initial value. Instead, the compact disk component reported in \citet{Masson+2016,Hennebelle+2016,WursterBate2019} do not present strong variability with the magnetic field intensity. There is a mild trend for the disk to be more massive and larger for weaker magnetization with high values of $\lambda$. 

To estimate the impact and consequences of diffusion we again consider two characteristic timescales relevant for the evolution of $B_\phi$ in the disk. On one hand, $B_\phi$ is as we saw earlier, generated by the differential rotation on a timescale $\tau_{\rm far}$ but on the other hand it is diffused vertically by ambipolar diffusion on a timescale $\tau_{\rm diff}$. For these simple estimates, we consider that ambipolar diffusion is the main diffusion process. We have
\begin{eqnarray}
\tau _{\rm far} &\simeq &{ B_\phi h \over B_z v_\phi}~, \nonumber\\
\tau _{\rm diff} &\simeq &{ 4 \pi  h^2 \over c^2  \eta_{\rm AD}  } {  B_z^2 + B_\phi ^2 \over B_z^2} \simeq { 4 \pi  h^2 \over c^2  \eta_{\rm AD}  }~.
\label{tau1}
\end{eqnarray}
Note that the time scales are estimated at the disk edge en route to deriving the disk size, hence the condition $B_\phi \ll B_z$ generally holds.

Again for a disk to form, it is likely that an equilibrium between these two time scales must occur in such a way that a stationary situation gets established, i.e. $\tau_{\rm far} \simeq \tau _{\rm diff}$. If the generation time of $B _\phi$ is faster than the diffusion time, then as mentioned before the disk quickly gets inflated. If the field diffuses too fast, magnetic braking will become continuously smaller. 

\begin{eqnarray}
{B_\phi \over  v_\phi} &\simeq& { 4 \pi h \over c^2 \eta_\mathrm{AD} }  B_z.
\label{far}
\end{eqnarray}

Second of all, for the disk to be in a stationary state, the braking and the rotation timescales must be of same order, $\tau _\mathrm{br} \simeq \tau_\mathrm{rot}$, yielding
\begin{eqnarray}
{ {B_\phi} \over v_\phi ^2 }  \simeq {2 h  \rho  \over r }  B_z^{-1}.
\label{br}
\end{eqnarray}

Combining  Eqs.~(\ref{far}) and~(\ref{br}, we thus get
\begin{eqnarray}
{2  \rho  \over r } v_\phi  \simeq { 4 \pi \over c^2 \eta_\mathrm{AD} }  B_z^2,
\label{vdhr}
\end{eqnarray}

Next, to obtain an estimate of the disk radius, we estimate the various quantities at the disk centrifugal radius location, i.e. at the disk-envelope boundary and the last quantity which is needed is the density.

In simple configurations and at an early time, typically at the epoch of the disk formation, we have
\begin{eqnarray}
\rho(r) = \delta { C_{\rm s}^2 \over 2 \pi G r^2} \left( 1 + {1 \over 2} \left( { v_\phi (r)\over C_{\rm s}} \right)^2 \right). 
\label{rho_exp}
\end{eqnarray}
Apart for $\delta$, which is a coefficient on the order of a few, the first term simply corresponds to the singular isothermal sphere (\ct{Shu1977}) while the second one is a correction that must be included when rotation is significant, particularly in the inner part of the envelope close to the disk edge, as discussed in \citet{Hennebelle+2004} (see their appendix and Fig.~2). Note that for very dynamical case, in particular for massive stars, $\delta$ may be  up to about 10 as shown in Fig.~3 of \citet{Hennebelle+2011}.  

In more complex situations, for example when the rotation axis and the magnetic field are misaligned but also generally, and for all configurations, at later times,  the disk connects to the envelope through an accretion shock. The typical velocity through the shock is the infall one, which is about $v_{\rm inf} \simeq \sqrt{G (M_*+M_ {\rm d}) / r_ {\rm d}} \simeq v_\phi$. Since the gas is isothermal, the density at the edge of the disk is simply: $\mathcal{M}^2 \rho_{\rm env}$. This expression is almost identical to the expression stated by Eq.~(\ref{rho_exp}).

Combining it with Eq.~\ref{vdhr}, we obtain
\begin{eqnarray}
{ \delta   G ^{1/2} (M_\mathrm{d} + M_*) ^{3/2} \over 2 \pi  r^{9/2} }  \simeq { 4 \pi \over c^2 \eta_\mathrm{AD} }  B_z^2,
\end{eqnarray}
where, for sake of simplicity, we have assumed $\rho \propto v_\phi^2$ in Eq.~(\ref{rho_exp}). This leads to an estimate for the disk radius that we call $r_{\rm ad}$, to stress the role of ambipolar diffusion:
\begin{eqnarray}
   r_{ad}  \simeq   \left(  {\delta G ^{1/2} c^2 \eta_\mathrm{AD}  \over 8 \pi^2 } \right)^{2/9} 
B_z^{-4/9}  (M_\mathrm{d} + M_*)^{1/3}.
\label{rad}
\end{eqnarray}

The poloidal magnetic field at the edge of the disk, $B_z$, needs also to be known. As discussed in \citet{Hennebelle+2016}, it tends also to be regulated and saturates around a value $\sim$0.1 $G$ if the initial field in the core is strong enough. If the field is initially weak, but also at later times when the collapsing envelope starts vanishing, $B_z$ may become much lower than this value. The mass of the star/disk system, $M_\mathrm{d} + M_*$, keeps increasing  as the envelope gets accreted. We adopt 0.1 $M_\odot$ as reference value.

With these values, Eq.~(\ref{rad}) can be rewritten:
\begin{eqnarray}
 r_{ad}  \simeq  18 \, {\rm AU}  \times
\delta ^{2/9 } \left( {   \eta_\mathrm{AD} \over 7.16 \times 10^{18} \, {\rm cm^2~s^{-1}}   } \right)^{2/9} 
\left( {B_z \over 0.1\,{\rm G} } \right) ^{-4/9}   \left( { M_\mathrm{d} + M_* \over 0.1 M _\odot} \right) ^{1/3}.
\label{radius}
\end{eqnarray}

This expression makes several interesting predictions. First the disk radius is initially on the order of 20~au and it then grows towards higher values as the mass, $M_*$, increases but also as $B_z$ diminishes. The disk radius increases, although sub-linearly, with the magnetic diffusivity. Finally, and surprisingly, it does not depend explicitly on the initial cloud rotation. This is possible obviously only if stationarity is possible regarding $B_\phi$ and therefore, rotation cannot be zero or very small, in which case Eq.~\ref{radius} is obviously invalid.

\subsubsection{Impact of initial conditions on disk formation}
\label{S.IC_impact}

Despite the numerical convenience, the axisymmetric set up, with uniform magnetic field and uniform cloud rotation, generally does not offer a realistic description of the protostellar collapse. Recent numerical efforts begin to introduce more complexities, such as different initial density distributions, misalignment between field and rotation, and turbulence. The results and the dependence on the various initial configurations sometimes depend on whether or not ideal or non-ideal MHD is applied.

\vspace{0.5cm}
\noindent $\bullet$ {\it The initial density profile}
\vspace{0.5cm}

\noindent An important question is whether for the same initial magnetic energy and angular momentum, the thermal energy of the core is influencing disk formation. Even more generally, to what extent is the initial density profile influencing the final protoplanetary disk?

This issue has been investigated in details by \citet{Machida+2014} and \citet{Machida+2016}, where 
several density profiles and thermal support are being considered. First, \citet{Machida+2014} compared uniform density clouds and clouds with a Bonnor-Ebert sphere like density profile. They found that while they had similar magnetization, namely $\lambda \sim 3$, the former tends to produce small disks while the latter produced much bigger ones. However, \citet{Machida+2014} stated that the initial magnetic field they used, is uniform. On the other-hand a critical Bonner-Ebert sphere has a density contrast between the edge and the center which is on the order of 10. This implies that the inner part of the clouds which has a Bonnor-Ebert mass density profile, has an effective $\lambda$ which is several times higher than 3. This is the most likely explanation of the large disk variations observed in the two types of calculations. 
Such a result is also confirmed in the study including AD (\ct{Lam+2019}), where the combination of a Bonner-Ebert sphere like density profile and a uniform magnetic field tend to facilitate disk formation. 

Second \citet{Machida+2016} performed calculations with uniform or Bonnor-Ebert sphere density profile but for various thermal supports, going from thermal to gravitational energy ratio, $\alpha$, of 0.2 to 0.7. They found that except in some of the most extreme cases where $\alpha > 0.5$, the disks have typically a radius of 15-20~au. There are a few extreme cases for which the disks are much smaller, which may be a consequence of a much longer collapse time scale leading to more time to transport angular momentum.



\vspace{0.5cm}
\noindent $\bullet$ {\it The impact of misalignment}
\vspace{0.5cm}

\noindent While the first calculations (e.g., \ct{Allen+2003,MellonLi2008,HennebelleFromang2008}) were performed assuming alignment between magnetic field and rotation axis, the effect of misalignment onto disk formation has been investigated in a series of papers. 

\citet{HennebelleCiardi2009} have run the first calculations and infer the surprising results that disks would more easily form in the misaligned configurations than in the aligned one. They propose that this could alleviate the magnetic braking catastrophe and indeed found that with $\lambda \simeq 5$, disk could form if magnetic field is sufficiently misaligned with respect to the rotation axis. This result has been confirmed by \citet{Joos+2012} and by \citet{Li+2013}, although these latter inferred slightly different critical $\lambda$ for disk formation. \citet{Joos+2012} have calculated the profile of angular momentum in the core envelope and clearly found that the angular momentum increases with $\alpha$, the angle between magnetic field and rotation axis. \citet{Gray+2018} also found that this is compatible with their studies. They performed calculations with an initial turbulent velocity field, which can be either random or for which the net angular momentum it contains has been artificially aligned with the magnetic field. They found no centrifugally supported cores in this latter case but did find disks in the former. 

The general interpretation comes from the geometry of the field lines (see \ct{Mouschovias1991,Joos+2012} for detailed discussions). Essentially as the torsional Alfv\'en waves generated by the disk propagate outwards, they sweep a certain amount of material in which they deposit angular momentum. The amount of material with which angular momentum is exchanged, depends on the field line geometry and so is the efficiency of the magnetic braking. If the magnetic field is aligned with the rotation axis and the magnetic field lines are uniform, then the amount of swept material is limited because the disk surface is small. In this case, the magnetic braking is more efficient if the field is misaligned. On the other-hand, if the field lines fan out, as it is the case in a collapsing cloud, then the amount of swept material may be higher in the aligned case.

A somewhat different result has been inferred by \citet{Matsumoto+2004} and \citet{Tsukamoto+2018}, who reported the angular momentum within the gas at very high density, i.e. $10^{-12}$ g cm$^{-3}$ or more and at relatively early time, that is to say at the moment when the protostar forms. They found that the angular momentum is higher in the aligned configuration than in the perpendicular one. Because of this higher angular momentum, they concluded that disks form more easily when the magnetic field is aligned. 

The reason of this apparent contradiction has not been investigated in detail yet. However, it should be noted that \citet{Matsumoto+2004} and \citet{Tsukamoto+2018} do not conduct the calculations up to the disk formation time. The gas in which their angular momentum is measured will not enter in the disk but rather form the star. One possibility is that at this early stage and for the aligned configuration, the magnetic field lines that permeate this very high density material, are relatively uniform because the material comes from a region located close from the cloud center. As time goes on, the material which reaches the center comes from further away and therefore the field lines open up more abruptly. A possible explanation is therefore that at early time, magnetic braking is more efficient in the perpendicular configuration but as the field lines get more and more stretched, magnetic field in the aligned configurations becomes more efficient.

Finally, several authors (\ct{Masson+2016,Tsukamoto+2018,WursterBate2019}) found that the impact of misalignment may be less pronounced when non-ideal MHD is treated. As explained previously, the polarity of the magnetic field can also affect the angular momentum distribution in the envelope-disk system via Hall effect. 

\vspace{0.5cm}
\noindent $\bullet$ {\it The role of turbulence}
\vspace{0.5cm}

\noindent The role of turbulence has also been recognized to be significant, at least in some circumstances, in collapsing cores. In the hydrodynamical case, \citet{Goodwin+2004a} and \citet{Goodwin+2004b} explored the influence of an initial turbulent velocity field on the collapse and fragmentation of 5 M$_\odot$ cores. They found that for an initial value of the turbulent over gravitational energy that is about 0.05, the number of objects formed is about 4-5 and the statistics of binaries are best reproduced. Although they did not study disks explicitly, the fragmentation occurs likely through massive disk instabilities.

In the context of magnetized cores, the influence of turbulence has also been explored in particular for its possible role in the transport of magnetic field and as a solution to the magnetic braking catastrophe. \citet{Santos-Lima+2012} first reported that while disk formation was prevented in the absence of turbulence, a small disk, could form when turbulence was present. This was a consequence of turbulent magnetic reconnection which leads to an effective transport of the magnetic flux and therefore a less efficient magnetic braking. Similarly, \citet{Seifried+2013} also found disk formation in the presence of turbulence. They however argued that this is not due to the turbulent diffusion but to the lack of ordered magnetic field. This disordered magnetic field, leads to inefficient magnetic braking. Note that this conclusion is different from the one reached by \citet{Gray+2018} who concluded that the dominant effect is the misalignment between rotation and magnetic field axis.

This has also been studied by \citet{Joos+2013} who performed a parameter study, considering several values of $\lambda$ and initial turbulent energy. They indeed confirmed that even when the turbulent energy is a few percent of the gravitational one, the magnetic flux in the inner part of the core was significantly reduced leading to the formation of a bigger disk, that would form sooner than in the case without turbulence.

The influence of turbulence has also been studied for non-ideal MHD collapse by \citet{Hennebelle+2016} and more recently by \citet{Hennebelle+2020}. They inferred that the disks which form in the runs with turbulence are not very different from the ones that formed in the runs without turbulence initially. The reason is probably that the diffusion of the magnetic field is dominated by the non-ideal MHD processes and turbulence is playing a secondary role in this respect.
Another possible effect is that turbulent motions may be partly quenched by the high value of the ion-neutral friction. \citet{Lam+2019} also performed turbulent collapse simulations with AD. They found that turbulence alone only allow the formation of disks at early times; to ensure the disk to survive into the later accretion phase, a relatively efficient AD (a few times the canonical value of \ct{Shu1991}) is also required. 

Recently, \citet{Verliat+2020} investigated the hydrodynamic collapse of motionless cloud (i.e. which has no turbulent and no rotation) but which presents density fluctuations. They found that RSDs also form because subsonic turbulence quickly develops and, angular momentum with respect to the moving star, is induced by the inertia forces. Therefore, the formation of initial disks can be a local process ($\sim$10$^3$~au scale; see \ct{Gaudel+2020}) and may not require organized rotation on the very large scale cloud ($\sim$10$^4$~au).

\section{Protostellar Disks around High Mass Stars}
\label{Chap.DiskHighMass}

\subsection{Observations of disk structures around massive protostars}
\label{S.ObsHighMass}


In comparison to low mass protostars, massive protostars are rare and they reside in star forming regions that are systematically farther away than low-mass star forming regions. Therefore, fully resolving disks around massive protostars remains challenging. Nonetheless, higher resolution and higher sensitivity observations in the last two decades have succeeded in imaging spatially resolved disks around high-mass YSOs, including some O-type stars (\ct{Cesaroni+2007,BeltranDeWit2016}). Observationally, it has been established that high-mass stars form via mass accretion through circumstellar disks at least up to $\sim$40~$M_\odot$, according to the most recent studies discussed below.

A prototypical high-mass disk is IRAS~20126$+$4104, located at a distance of 1.64~kpc and studied in detail with the PdBI and the SMA (\ct{Cesaroni+2005,Cesaroni+2014,Xu+2012,Palau+2017}). Sub-arcsecond resolution observations of thermal molecular line emission from species such as methyl cyanide (CH$_3$CN) resolved a compact structure rotating around a central mass of $\sim$10~$M_\odot$. At these resolutions, the observed disks have typical sizes of about 1000~au, while some high resolution maps show a few 100~au scale disks (\ct{BeltranDeWit2016}). At higher resolutions, about 1~milli-arcsecond using VLBI networks (\ct{Bartkiewicz+2009,Fujisawa+2014}), the 6.7~GHz methanol masers are used to reveal dynamical structures around high-mass young stellar objects (YSOs) as they are exclusively excited by strong far-infrared radiation in high-mass and intermediate-mass YSOs. Some methanol maser sources show circular, elliptical or linear structures with velocity gradients consistent with rotating disks, e.g., Cepheus~A HW2 and NGC7538~IRS1 (\ct{Sugiyama+2014,MoscadelliGoddi2014}). Near-infrared interferometers such as the VLTI reveal hot circumstellar disk candidates at scales $<$100~au, thanks to their achieved milli-arcsecond resolution observations (\ct{Kraus+2010,Boley+2013}).

Earlier reviews (\ct{Cesaroni+2007,BeltranDeWit2016}) have summarized the general properties of disks around high-mass YSOs with masses up to 25--30~$M_\odot$ and found that these disks are consistent with scaled-up versions of disks around low-mass YSOs, though with larger sizes, more massive and higher mass infall/accretion rates. In contrast, it is unclear whether O-type stars with masses larger than 40~$M_{\odot}$ can be formed in a similar way, since these objects are usually associated with more massive ($>$100~$M_\odot$) and larger ($>$10$^{4}$~au) rotating structures called toroids (\ct{Cesaroni+2007,BeltranDeWit2016}). These massive toroids are gravitationally unstable and prone to fragmentation and collapse.

The advent of ALMA and its longest baselines (up to 16~km) have enabled sensitive studies at resolutions of a few 10~milli-arcseonds, corresponding to 100~au or better spatial scales. Even shortly after the publication of the previous review by \citet{BeltranDeWit2016}, there has been significant progress in observational results for new disk candidates and more detailed studies of already known disk sources.

%
\subsubsection{Massive disk and infalling-rotating envelope structures}

ALMA observations of low-mass disks like L1527 in Taurus and IRAS~16293$-$2422 in Ophiuchus (see \S~\ref{S.ObsLowMass}) have revealed that chemical composition changes drastically within the scales of the disk and envelope, enabling the study of these different components. 
Similarly, in IRAS~16293$-$2422, a YSO rich in complex organic molecules (COMs; \ct{Bottinelli+2004}), the envelope is seen in sulfur-bearing species such as OCS, CS and H$_2$CS (\ct{Oya+2016}), while the centrifugal barrier and the Keplerian disks are detected in COMs and H$_2$CS, respectively. These results demonstrate that high resolution images of various molecular lines are useful to differentiate dynamical structures essential in the mass accretion process.

Similar to the low-mass case, recent ALMA observations of high-mass YSOs have resolved the disk and infalling envelope when observing different species at high-enough resolution (\ct{Csengeri+2018, Maud+2018, Motogi+2019, Zhang+2019}). For instance, the massive protostellar source G339.88$-$1.26 shows chemical differentiation in species such as SiO, SO$_2$, H$_2$S, CH$_3$OH, and H$_2$CO (\ct{Zhang+2019}), with CH$_3$OH and H$_2$CO lines likely tracing the infalling-rotating envelope, and the SO$_2$ and H$_2$S lines being enhanced around the centrifugal barrier. Furthermore, SiO is detected in both the Keplerian disk and the outflow which is likely driven by a disk wind. Because of the difference in luminosity of the central protostar and hence, in the surrounding gas, the chemical composition of the different structures may be significantly different compared to those in low-mass objects. The size of the centrifugal barrier, about 500~au in G339.88$-$1.26, is significantly larger than those in low-mass objects, $\sim$200~au, and the resultant specific angular momentum is a factor 10 larger (\ct{Zhang+2019}). This chemical differentiation is also seen in the envelope-disk system G328.2551$-$0.5321, where torsionally excited CH$_3$OH and vibrationally excited HC$_3$N lines are emitted in the innermost region of the envelope, possibly coincident with the centrifugal barrier and the compact Keplerian disk, respectively (\ct{Csengeri+2018}). Such an ordered envelope-disk structure suggests that the accretion process in high-mass objects like G339.88$-$1.26 is consistent with a scaled-up version of the accretion process in low-mass protostars.

Mass infall rates of $10^{-4}$--$10^{-3}$~$M_\odot$~yr$^{-1}$ have been measured towards the high-mass YSO G351.77$-$0.54 via the blue-shifted absorption lines at high-frequencies in ALMA band~9 observations (\ct{Beuther+2017}). The disk-jet system in the O-type YSO G23.01$-$0.41 shows sub-Keplerian rotation and infalling motions in the disk with an accretion rate of $6\times10^{-4}$~$M_\odot$~yr$^{-1}$ (\ct{Sanna+2019}). The face-on disk source G353.273$+$0.641, associated with 6.7~GHz methanol masers, shows a velocity field consistent with infall motions along parabolic streamlines falling onto the equatorial plane of the disk (\ct{Motogi+2017}). All results to date are consistent with accretion rates $>10^{-4}$~$M_\odot$~yr$^{-1}$, a few orders of magnitude larger compared to their low-mass siblings.

\subsubsection{Keplerian rotating disks}

\begin{figure}[ht]
\centerline{\includegraphics[width=0.65\textwidth, angle=-90]{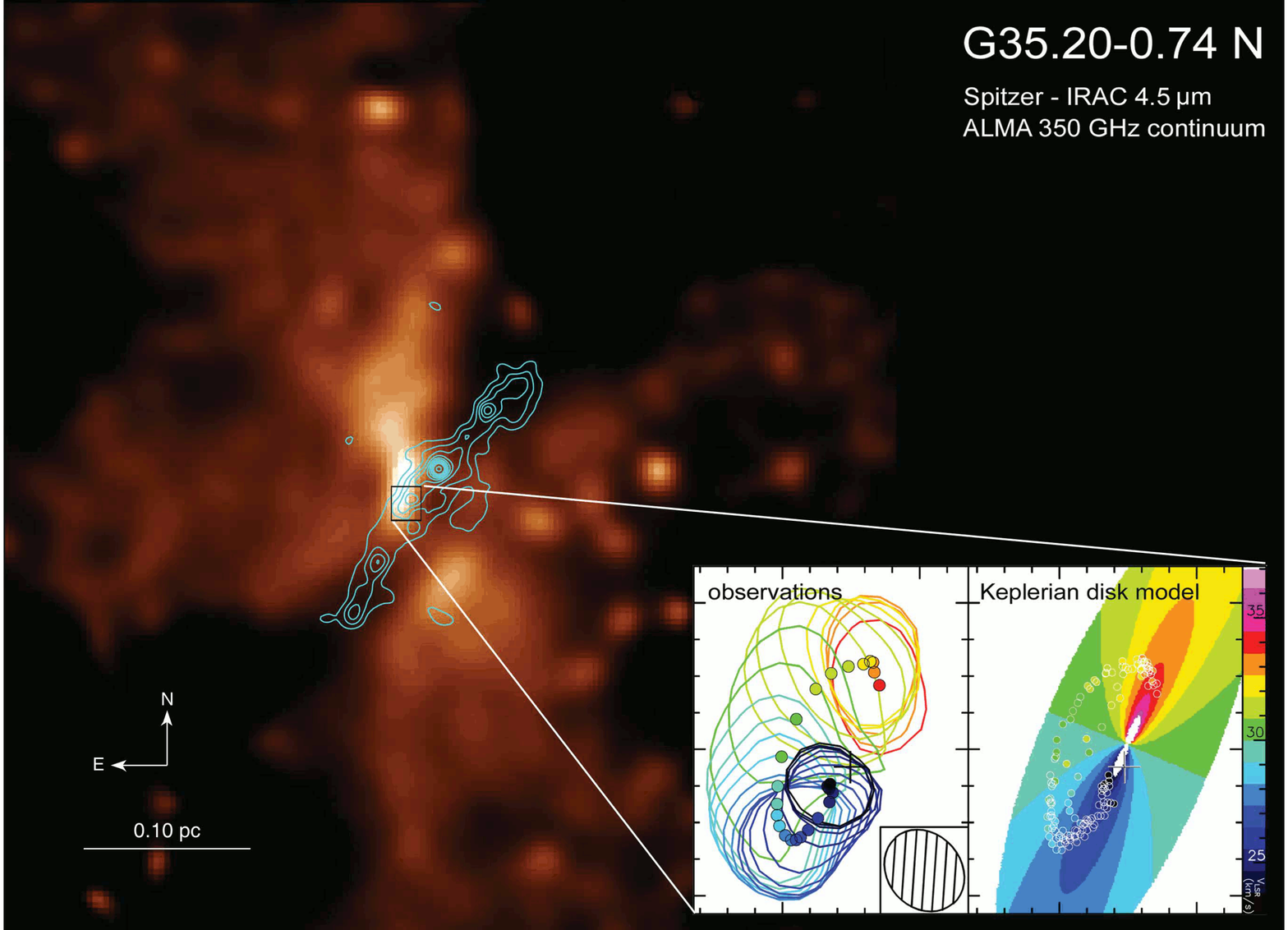}}
\caption{ALMA observations of continuum and molecular line emission with 0.4\arcsec\ resolution towards the high-mass star forming region G35.20$-$0.74~N. Two dense cores are detected in typical hot-core tracers, such as CH$_3$CN, and they reveal velocity gradients. In one of these cores, the velocity field can be fitted with an almost edge-on Keplerian disk rotating about a central mass of 18~$M_\odot$. Figure from \citet{Sanchez-Monge+2013}.}
\label{Fig:G3520N}
\end{figure}

Recent examples of rotating disks around high-mass YSOs are G35.20$-$0.74N at 2.19~kpc (Fig.~\ref{Fig:G3520N}; \ct{Sanchez-Monge+2013,Sanchez-Monge+2014}), G17.64$+$0.16 at 2.2~kpc (\ct{Maud+2019}), IRAS~16547$-$4247 at 2.9~kpc (\ct{Zapata+2015,Zapata+2019}), G35.03$+$0.35 at 3.2~kpc (\ct{Beltran+2014}), G11.92$-$0.61 at 3.37~kpc (\ct{Ilee+2018}), G16.59$-$0.05 at 3.6~kpc (\ct{Moscadelli+2019}), and AFGL4176 at 4.2~kpc (\ct{Johnston+2015}), ordered in distance from the nearest to the farthest disks. They exhibit probably Keplerian rotation curve observed in CH$_3$CN and CH$_3$OH lines including high K-ladder lines, torsionally excited transitions, and isotopologues, which preferentially trace hotter and denser regions.

\begin{figure}[ht]
\centerline{\includegraphics[width=0.75\textwidth]{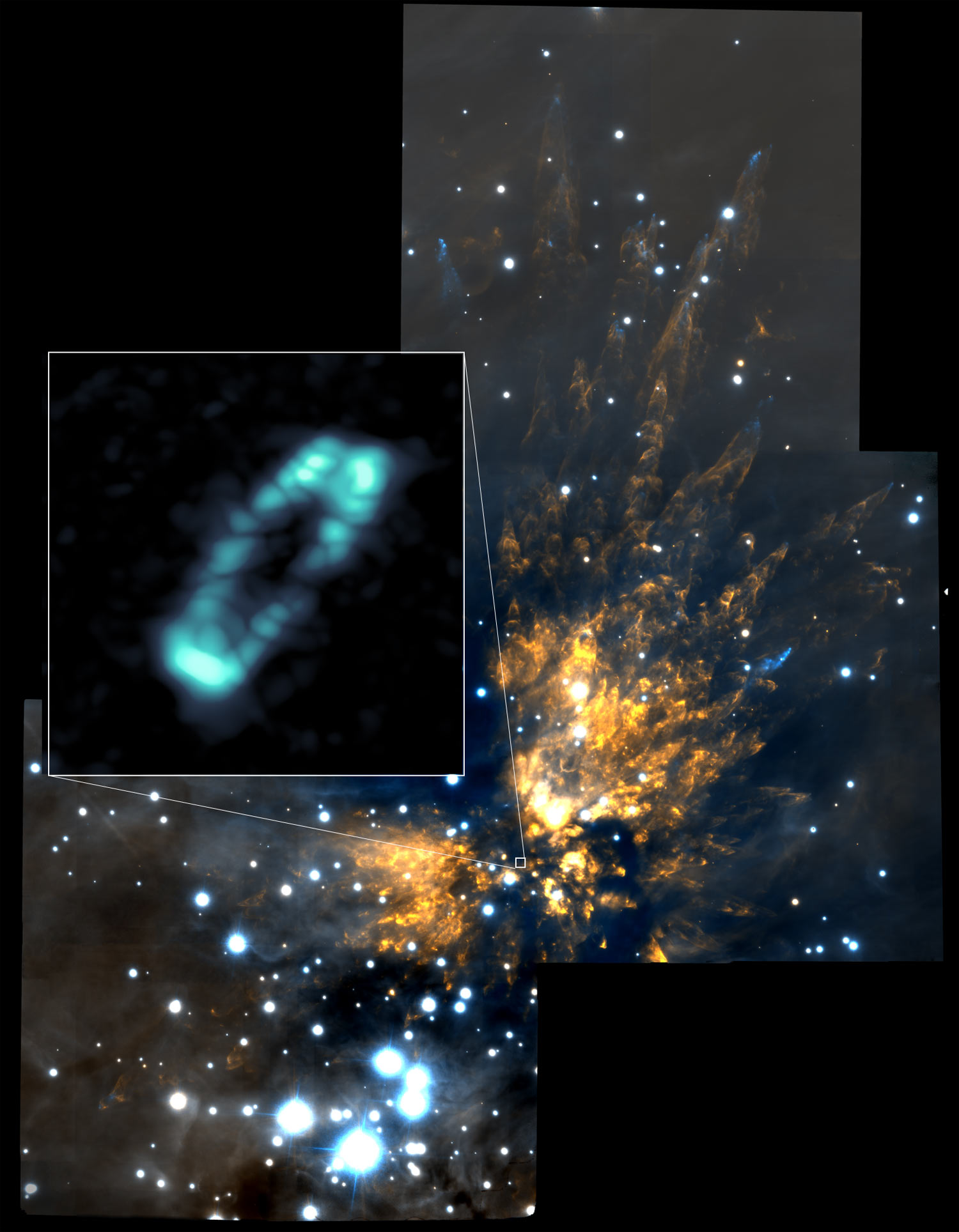}}
\caption{ALMA image of the salty disks surrounding the young, massive star Orion Source I (blue ring). It is shown in relation to the explosive region in the Orion Molecular Cloud 1 region. Image credit: ALMA, NRAO, ESO, NAOJ, AUI, NSF, Gemini Observatory, AURA.}
\label{Fig:OrionSrcI}
\end{figure}

Although it is still difficult to accurately determine the rotation velocity curve as a function of radius, $v \propto r^{-\alpha}$, high excitation molecular lines present more clear evidences of Keplerian(-like) rotation disks consistent with $\alpha \sim$0.5. 
The best example is found towards the nearest high-mass YSO Orion~Source~I, located at only 420~pc (Fig.~\ref{Fig:OrionSrcI}; \ct{Ginsburg+2018,Ginsburg+2019}) and consistent with a central mass of 15~$M_{\odot}$, slightly higher than the results from previous lower resolution observations: 5--7~$M_{\odot}$ (\ct{Kim+2008,Hirota+2014,Plambeck+2016}).

\subsection{Numerical models for disk formation in high mass stars}
\label{S.TheoryHighMass}

In the context of individual star and disk formation, the basic physical processes and relations of disk formation, disk growth, and disk evolution is expected to be similar for low-mass and high-mass stars.
Nonetheless, one aspect can be highlighted as an important difference: low-mass stars and high-mass stars are expected to form from different initial conditions, including the typical mass reservoirs in terms of mean density, the amount of thermal support against gravity, the large-scale angular momentum distribution, the turbulence properties of the surrounding, stellar radiation, and the potential impact of a forming star cluster.

In the context of high-mass star formation, disk formation and evolution has been studied in a variety of simulations with and without magnetic fields and with and without the effects of radiation transport.
With respect to the questions of angular momentum transport from clouds to disks, disk formation and disk growth, a variety of MHD simulations have been performed (\ct{BanerjeePudritz2007,Hennebelle+2011,Peters+2011, Seifried+2011,Commercon+2011,Myers+2013,Matsushita+2017,KolliganKuiper2018}), although the majority of those focus on the impact of magnetic fields on cloud collapse and cloud fragmentation. 
The conclusion of those investigations is in line with the results reported for the low-mass star formation case, namely that ideal MHD simulations yield the formation of magnetic-pressure-dominated pseudo-disks (\ct{Hennebelle+2011,Peters+2011,Commercon+2011}).  Magnetic braking can be circumvented and disk growth enabled by either misalignment between the rotation and magnetic field axis, possibly caused by turbulence (\ct{Seifried+2012}), or non-ideal MHD effects (\ct{Matsushita+2017,KolliganKuiper2018}). 
These processes show a dependence on the mass-to-flux ratio of the initial mass reservoir (\ct{Seifried+2011}) and still a strong dependence on spatial resolution (\ct{Hennebelle+2011,Myers+2013,KolliganKuiper2018}), i.e.~the MHD physics are not fully converged yet. 
In the case of high-mass star formation the convergence aspect might be more severe as in the low-mass case due to stronger numerical diffusion at high density. From these MHD simulations, only the most recent studies have enabled sub-au resolution and observed the launching of magneto-centrifugally-driven jets from the disk surfaces (\ct{Matsushita+2017,KolliganKuiper2018}).

Disks around very massive stars are further impacted by the strong radiation from their host stars. Although the high optical depth of the disks in fact enables disk accretion above the so-called radiation pressure barrier (\ct{Kuiper+2010}) at first, later in evolution, the strong radiative impact limits the accretion flow toward the host star in several ways:
On large scales, continuum radiation forces are predicted to determine the lifetime of accretion disks around these super-Eddington stars (\ct{KuiperHosokawa2018}). On small scales, the disk-to-star accretion flow has to overcome the energetic impact of UV-line-driven radiation forces (\ct{Kee+2018a,Kee+2018b,KeeKuiper2019}), which might be the reason for an intrinsic upper mass limit of stars by accretion (\ct{KeeKuiper2019}).


\section{Gravitational Instability in Massive Protostellar Disks}
\label{Chap.GIFrag}

Once a small disk is formed in the early phase of the star formation process, the disk grows via accretion from the envelope during the main accretion phase. Since the gas from the outer envelope has a larger angular momentum content, the infalling gas tends to arrive 
in the outer region of the disk, causing the disk to grow in size as the star formation process precedes. If the mass and angular momentum within the disk are not transported by magnetic processes sufficiently faster than accretion, the disk gets considerably massive compared to the central protostar and eventually becomes gravitationally unstable. This is likely to occur when the non-ideal MHD processes work efficiently, but it can realize even in the ideal MHD limit depending on the initial mass and angular momentum distributions.

The gravitational instability is well characterized with the Toomre's Q parameter (\ct{Toomre1964}): 
\begin{equation}
Q = \frac{\kappa c_s}{\pi G \Sigma},
\end{equation}
where $\kappa$ is the epicyclic frequency ($\kappa = \Omega$ if the rotation profile is Keplerian), $c_s$ the sound speed of the disk, and $\Sigma$ is the surface density, respectively. When this value is lower than unity, it means that the disk is gravitationally unstable against axisymmetric perturbation. Even when it is higher than unity, it can still be unstable against non-axisymmetric perturbation as long as $Q$ value is not too large. While the exact threshold for the instability depends on the disk structure, typically the disk remains unstable if Q is as small as a few (\ct{Takahashi+2016}). In this regime, the disk mass can be as high as 30-40\% of the central protostar (\ct{Rosen+2019}). Spiral arms are formed by the instability, and transport angular momenta by self-gravitational torques. 

The fate of the disk is determined by competition between the accretion which makes the disk unstable and the angular momentum and mass transport which makes the disk stable. When the accretion is modest, the disk can maintain a self-regulated state where the two processes are balanced. Because the disk has a shearing rotation profile close to the Keplerian rotation, material spiral arms cannot be a steady structure. They are wound up by the shear rotation and disappear in a few orbits. However, the accretion makes the disk unstable again and spiral arms form recurrently (e.g., \ct{Tomida+2017}). On the other hand, when the accretion rate is so high that the disk cannot maintain the balance, the gravitational instability grows and the disk fragments into binaries or multiples. While binaries can be also formed by other mechanisms including turbulent fragmentation in the core scale (e.g., \ct{Rosen+2019}), the disk fragmentation is still a plausible process, and these processes are not exclusive but can work together. 

Early analytical study of gravitational instability (e.g., \ct{PapaloizouSavonije1991}) was followed by hydrodynamics simulations (e.g., \ct{Kratter+2010}), in which large spiral arms and multiple systems easily form in massive accreting disks. Recent non-ideal MHD simulations for low mass star formation (\ct{Zhao+2018a,WursterBate2019}) demonstrated that faster rotation and/or lower magnetization ($\lambda \gtrsim 5$) of the initial core facilitate the development of spiral arms and companion fragments around the central disk. 
Protostellar disks around high-mass stars tend to be more massive and more susceptible to gravitationally instability and fragmentation than those around low-mass protostars. However, it remains an open question as to what exact extent disk fragmentation contributes to the multiplicity properties of high-mass star forming regions. Disk formation and fragmentation around forming massive protostars has been studied in only a few three-dimensional simulation studies which include radiation transport to properly render the thermodynamics of the (un)stable disk (\ct{Krumholz+2007,Krumholz+2009,Kuiper+2011,Rosen+2016,Klassen+2016,Meyer+2017,Meyer+2018, Rosen+2019}).

However, disk fragmentation in numerical studies is dependent on numerical resolution. Usually, increasing the resolution promotes fragmentation. While \citet{Kuiper+2011} only yields to the formation of spiral arms, these spiral arms fragment in the higher resolution versions (\ct{Meyer+2017,Meyer+2018}).
Often, due to limited resolution (e.g., 10--20~au in high mass simulations), sink particles are used to follow the evolution of individual stellar components. However, the conditions for fragmentation also depend on the sink particle creation, accretion and merging criteria (e.g., \ct{Krumholz+2004,Federrath+2010}). Moreover, simulations forming fragments often show migration of fragments towards the primary star due to gravitational torques, causing accretion luminosity bursts (\ct{Meyer+2017,Zhao+2018a}) and the formation of spectroscopic binaries (\ct{Meyer+2018}).

Despite the common phenomenon of gravitational instability and fragmentation, observations have only recently start to resolve the disk sub-structures such as spiral and fragments in around protostars (e.g., \ct{Ilee+2018, Zapata+2019}). 

\subsection{Circumbinary, circum-multiple disks, and spiral structures in low mass protostars}
\label{S.GILowMassObs}

\begin{figure}[ht]
\centerline{\includegraphics[width=0.75\textwidth]{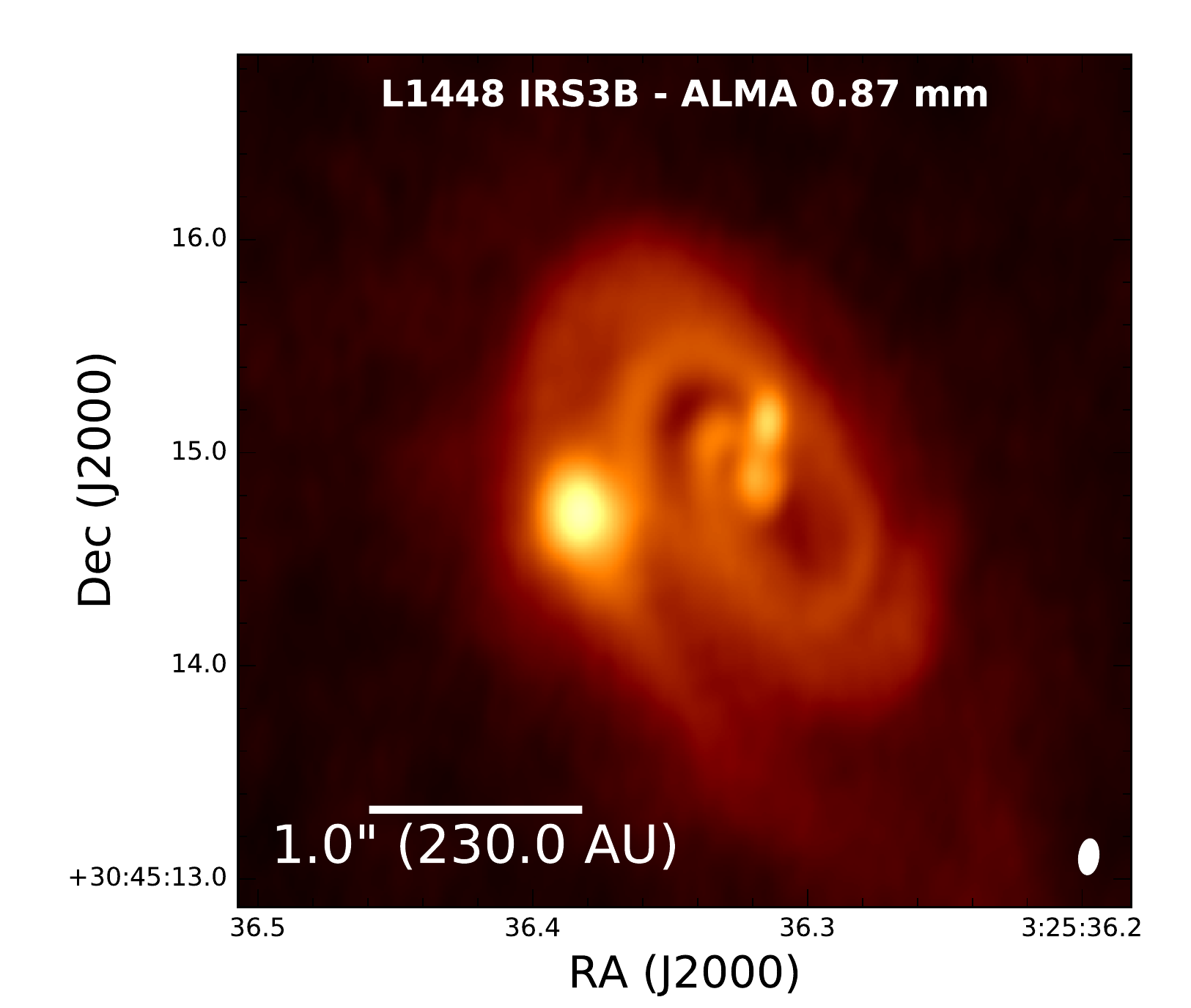}}
\caption{ALMA 0.87~mm image of the Class 0 L1448 IRS3B from ALMA with $\sim$0.15\arcsec\ (45~au) resolution. The kinematic center of mass is near center of the image where there is a small depression of emission between two clumps. The three structure of the clumpy emission in the disk point to possibly three protostars forming within the circum-multiple disk.}
\label{Fig:L1448}
\end{figure}

The observed faction of binary and multiple systems in low mass protostars are already high.
A complete survey for multiplicity toward Perseus protostars down to $\sim$25~au showed that at least $\sim$28\% of Class 0 protostar systems have a companion separated by less than 1000~au (\ct{Tobin+2016a}). Furthermore, the majority of these multiple systems with separations less than 1000~au have separations less than 200~au, scales approaching the sizes of protostellar disks. Sensitive and high dynamic range sub-arcsecond resolution observations from ALMA revealed a few young circumbinary structures. Circum-multiple disks have been detected toward a few protostars in Perseus (\ct{Tobin+2018}), and an exemplar of this sample is L1448 IRS3B (Fig.~\ref{Fig:L1448}), where there is a protostar in the outer disk that has brighter peak submillimeter intensity than the circumstellar disk(s) toward the central protostar(s). The spiral structure in this disk and the formation of the companion are consistent with being triggered by gravitational instability, with the estimated Toomre $Q$ parameter of around unity between 150--320~au (\ct{Tobin+2016b}). However, recent CALYPSO survey (\ct{Maret+2020}) shows that the gas kinematics around L1448 IRAS3B do not follow a Keplerian rotation curve, cautioning that the circum-multiple disk may differ from the well-defined Keplerian disks around single stars. The spiral structures may also be viewed as rotating structures that connects individual circumstellar disks (\ct{Maury+2019}) and the infalling envelope. Accordingly, the companion may be formed by rapidly piling up infalling materials in the spiral structures, as discussed by \citet{Zhao+2018b}.

Regarding more evolved Class I systems, L1551~NE was shown to harbor a R$\sim$300~au circumbinary disk that appears to be largely the remnant of the binary star formation process in that system (\ct{Takakuwa+2017}).
Finally another extraordinary circumbinary disk is found toward the Class I intermediate mass protostar [BHB2007] 11, where spiral arms are apparently excited by the binary interaction of the protostars separated by $\sim$30~au and have circumstellar disks that are 2-3~au in radius (\ct{Alves+2019}).

Furthermore, recent ALMA collaborations (e.g., DSHARP) also reveal spiral structures in a few evolved Class II systems (Elias 27, IM Lup, WaOph 6; \ct{Perez+2016,Huang+2018}), with $m=2$ symmetry.  While it is not easy to measure the mass of the disk because of the large uncertainty in the dust opacity, such spiral substructures are clear evidence of gravitational instability in operation in these evolved stages.

%
\subsection{Gravitational instabilities and binary formation in high mass sources}
\label{S.GIHighMassObs}
In massive protostellar sources, earlier results showed that most Keplerian disk candidates are stable with $Q>1$ (\ct{BeltranDeWit2016}), in which the temperature of the disk are probed by observations of rotational lines of species such as CH$_3$CN and CH$_3$OH, and the surface density is estimated from the dust continuum emission.
The more recent studies for G351.77$-$0.54 and G17.64$+$0.16 are consistent with this scenario, both having disks that appear stable against collapse and fragmentation (\ct{Beuther+2017,Maud+2019}). The $Q$ parameter for G17.64$+$0.16 reaches its lowest values of 2--4 towards the ring-like structure seen in dust emission.

Despite these results, the majority of high mass disk candidates are thought to be unstable (\ct{BeltranDeWit2016}). A more recent study towards the spatially resolved face-on disk in G353.273$+$0.641, is consistent with the disk being gravitationally unstable with average Toomre $Q$ values on the range 1--2 and 0.4 at its minimum (\ct{Motogi+2019}). This is consistent with the existence of non-axisymmetric structures in the disk, indicative of a spiral arm and condensations with masses of 0.04--0.09~$M_\odot$. These condensations formed via disk fragmentation could eventually form a binary companion as seen in a low-mass multiple system (\ct{Tobin+2016b}) or accrete onto the central massive YSO resulting in accretion burst events (\ct{CarattiOGaratti+2017,Hunter+2017}). The high-mass YSO G11.92$-$0.61 hosts a binary system with a extremely large mass ratio. The main core has a mass of 40~$M_\odot$ and the companion, which may have formed via fragmentation of the disk, is only $<0.6$~$M_\odot$ (\ct{Ilee+2018}). Another binary system is found in IRAS~16547$-$4247, which shows an asymmetric structure possibly caused by fragmentation within the disk (\ct{Zapata+2019}). The disk associated with the high-mass YSO G35.20$-$0.74N is likely a circumbinary disk rotating about two small cores detected at radio wavelengths (\ct{Sanchez-Monge+2013,Beltran+2016}). Disk instability is also reported for the disk in W3(H$_2$O) where the Toomre $Q$ analysis suggests that the unstable disk may form a binary system via its fragmentation (\ct{Ahmadi+2018}). Although the statistics, in terms of objects analyzed, are still limited, and the determination of the Toomre $Q$ parameter highly depends on the uncertainties in the temperature and surface density determinations, these new results favor a scenario where disks around high-mass YSO are highly unstable and prone to fragment and form binary systems.

%
\subsection{Disk-mediated accretion bursts}
\label{S.GIBurstObs}

The fragmentation of the disk may cause episodic mass accretion burst events 
Similar to the low-mass FU-Ori and EX-Lup objects, a sudden increase in the flux density at infrared and radio wavelengths has been reported in the high-mass YSOs NGC6334I-MM1 (\ct{Hunter+2017,Hunter+2018,Brogan+2018}) and S255IR~NIRS3 (\ct{CarattiOGaratti+2017,Cesaroni+2018,Liu+2018}). The bolometric luminosity of S255IR~NIRS3 increased up to $1.3\times10^{5}$~$L_\odot$ which corresponds to a mass accretion rate of $5\times10^{-3}$~$M_\odot$~yr$^{-1}$. In the case of NGC6334I-MM1, the luminosity has been estimated to increase by a factor of 70 (up to $4.2\times10^{4}$~$L_\odot$) compared to the luminosity in the pre-flare or pre-burst phase (\ct{Brogan+2016,Hunter+2017}). These bursts are associated with flares in the emission of the 6.7~GHs methanol masers (\ct{Fujisawa+2015,MacLeod+2018,Moscadelli+2017,Szymczak+2018}). Thus, maser flares constitute an ideal probe to identify the initial phase of accretion bursts. Recent discoveries of other methanol maser flares have triggered intensive observational studies of possible mass accretion events in G358.93$-$0.03 (\ct{Sugiyama+2019, Brogan+2019,Burns+2020}). Statistical studies of accretion bursts in high-mass YSOs are necessary to compare to those found in the low-mass FU-Ori and EX-Lup cases. 


\section{Outflow and Jet Launching}
\label{Chap.Outflow}

The formation of disks are often accompanied by the launching of jets and outflows, which transport away the angular momentum from the accreting material. Recent high-resolution observations start to distinguish between the collimated jets and the wide-open angle outflows, which may have different origins.

When a rotating magnetized cloud core collapses, the gas spins up as it collapses and magnetic field lines get twisted. If magnetic fields are very weak, the fields are passively wound up by rotation. The toroidal fields are amplified and the gas near the pole gets accelerated by the magnetic pressure and forms outflows. In this case, because the toroidal magnetic fields confine the gas, the outflows, or jets, are well collimated. On the other hand, when magnetic fields are relatively strong, then the magneto-centrifugal force accelerates the gas in the disk outward along the field lines. In this case, the outflows have wide opening angle (\ct{BlandfordPayne1982}). Because the outflowing gas can carry the angular momentum away as the bulk angular momentum, the outflows contribute to the angular momentum transport considerably. Both processes can accelerate the outflows as fast as the rotation velocity at the launching point, because ultimately the source of the energy is the gravitational energy.

Recent MHD simulations can explain both outflow processes as results of interaction between magnetic fields and rotating gas. The non-ideal MHD simulations of \citet{Machida+2006} demonstrated that, in molecular cloud cores with typical rotation and magnetic fields, the wide-opening magneto-centrifugal outflows are driven from the first core and disk scale ($r \sim 10 $~au). In the central region of the cloud, the magnetic fields are dissipated by non-ideal MHD effects, and the weak magnetic fields drive the magnetic pressure mode in the vicinity of the central star. Because the latter is driven from the region closer to the star, the gravitational potential is deeper and the resulting outflows are much faster. This scenario naturally explains the two-component structures of protostellar outflows with the slow, wide-opening molecular outflows and fast, well-collimated jets.

In high-mass star formation processes, the central protostar produces strong radiation while the gas in the envelope is still accreting. In spherically symmetric configuration, this radiation can prohibit accretion beyond a certain mass (\ct{WolfireCassinelli1986}). However, in realistic situation, accretion through a circumstellar disk is possible, and radiation feedback does not completely halt the accretion. Still, the strong radiation force from the central protostar can drive outflows and regulate the star formation efficiency to 50 - 70\% (\ct{YorkeSonnhalter2002,Krumholz+2009,Kuiper+2015,Rosen+2016}). Strong UV radiation can ionize the envelope and produce an HII region of $T\sim 10^4 {\rm K}$, which can also suppress the accretion.

While the radiation processes work efficiently only in high mass star formation, the magnetic processes work in any mass regime as long as sufficient magnetic fields exist. Recent MHD simulations showed that the outflows launched by magnetic fields can carry away a significant fraction of accreting gas, yielding star formation efficiency of 20 - 50\% (\ct{MachidaHosokawa2013}), which can significantly contribute to regulate the final mass of formed stars. As the ejection rate by the MHD outflow is roughly proportional to the accretion rate, this mechanism can work both in the low mass and high mass regimes (\ct{Matsushita+2018}), and some numerical and semi-analytic studies even suggest that the magnetically-driven outflows dominate the radiation-driven outflows (\ct{Tanaka+2017}). These results infer that high mass star formation can be qualitatively similar to low mass star formation. Still, it is of crucial importance to study the combined effects of various physical processes using realistic simulations.

%
\subsection{Observations of outflow launching and rotation}

Recent high-resolution interferometry has enabled observations to reveal the detailed structures and kinematics of jets and outflows around both low- and high- mass sources. 

\citet{Bjerkeli+2016} showed that the molecular outflows in TMC1A are carrying significant angular momentum, and driven from the $\sim 10 $~au scale. Obvious offset of outflow launching site relative to the central stellar object is also observed in \citet{Alves+2017} of the Class I source BHB07-11. These wide angle outflows are better explained by the disk wind models (\ct{PelletierPudritz1992}), rather than X-wind (\ct{Shu2000}) or stellar wind models. In comparison, \citet{Lee+2017} detected knots of SiO along the collimated jet of the Class 0 source HH212. Velocity gradient across these knots perpendicular to the jet is also detected with 7$\sigma$ level, from which a mean specific angular momentum of $10.2\pm1.0$~au~km~s$^{-1}$ is measured in the jet. Hence, the jet launching radius can be derived as small as 0.05~au, indicating a possible X-wind origin for the collimated jet.

Similar observations are performed for a high mass protostar. 
Possible evidence of an outflow driven by a disk wind has been proposed for Orion~Source~I based on the three-dimensional velocity field measured with maser proper motions (\ct{Matthews+2010,Greenhill+2013}). More recently, ALMA observations of a high excitation Si$^{18}$O line have revealed a velocity gradient in the outflow consistent with the rotation of the disk (\ct{Hirota+2014,Hirota+2017}). The position-velocity diagram suggests an enclosed mass of $8.7\pm0.6$~$M_\odot$ and a centrifugal radius, which is close to the outflow launching radius, of 21--47~au. These results suggest that the outflow is most likely driven by a magneto-centrifugal disk wind (\ct{BlandfordPayne1982,Matsushita+2017}). The disk-outflow system in Orion~Source~I may have a similar origin to what is found in low-mass sources, with a larger launching radius possibly due to a larger initial specific angular momentum (about one order of magnitude larger; \ct{Motogi+2017,Csengeri+2018}). The disk and outflow launching regions have been investigated by observing vibrationally excited lines of H$_2$O, SiO and metal compounds (\ct{Hirota+2012,Hirota+2014,Hirota+2017,Kim+2019,Ginsburg+2018,Ginsburg+2019,Tachibana+2019}). The strong radiation of the central object and/or shocks associated with the outflow/wind launching regions enhance the gas-phase abundances of these species and excite them to high excitation energy levels. One of such transitions is the H$_2$O line at 232~GHz, $\nu_{2}$=1,  $5_{5, 0}$-$6_{4, 3}$, that has an excitation energy level of 3500~K (\ct{Hirota+2012, Ginsburg+2018}). This line is also detected in other high-mass YSOs such as G17.64$+$0.16 (\ct{Maud+2019}) and G351.77$-$0.54 (\ct{Beuther+2019}) using ALMA. Future higher resolution observations of other sources will test whether the case of Orion~Source~I is common to high-mass YSOs.

Nevertheless, this resemblance in the outflows between the low- and high- mass stars may be another indication that magnetic processes are dominant in outflow driving across broad mass range, and high-mass star formation can be actually a scale-up version of low-mass star formation.

\section{Discussion}
\label{Chap.Discuss}

\subsection{Observational caveats in detecting protostellar Disks}

As mentioned in \S~\ref{S.ObsLowMass}, the envelope and disk emission in protostellar systems is entangled, and it is observationally difficult to separate continuum and molecular line emission into envelope and disk components. Moreover, even in cases where the Keplerian radius of the disk can be discerned from the rotation curve (Fig.~\ref{Fig:L1527}), there is still ambiguity as to whether the disk extends beyond this radius or not. 
Thus, a powerful tool in revealing the properties of embedded protostellar disks is radiative transfer modeling. While models typically need to adopt asymmetric density structures to limit the computational complexity and parameter space, they can reasonably approximate the structure on less than 1000~au scales and the most significant deviation from symmetry is likely on large scales (e.g., \ct{Tobin+2010}). Modeling can be simple where intensity profiles are fit to the visibility amplitudes without costly radiative transfer calculations to estimate the disk properties (\ct{Maury+2019}). Alternatively, Markov Chain Monte Carlo (MCMC) modeling using full radiative transfer (radiative equilibrium) to
model both the spectral energy distributions and the continuum visibilities from the envelope and disk can be employed (e.g., \ct{Chiang+2012,Sheehan+2017b}). Both approaches have merit, but the full radiative transfer models present the most realistic representation of a
protostellar system with spatially varying temperature profiles and continuum opacity taken into account. Furthermore, if radiative transfer modeling is conducted at multiple wavelengths, the continuum opacity of the disk can be determined more robustly to better constrain the disk masses, but the dust opacities must be known a priori.

In addition to continuum model, spectral line modeling of the disk and envelope systems can also be carried out (e.g, \ct{Aso+2017}). This enables the contributions of both the disk and the envelope to be quantitatively fit to the data to better characterize the likely radius of the disk from the molecular line emission. Moreover, if carried out in a MCMC approach confidence intervals for the Keplerian disk radii can be ascertained. Furthermore, since both the disk and envelope are being modeled simultaneously, the analysis of the rotation profile of the disk and envelope are not limited to just using the highest velocity emission, where the signal-to-noise is often the lowest. Therefore, spectral line modeling can more effectively utilize the full spatial dynamic range of the data to characterize the rotation of the disk and envelope as a whole and derive accurate protostellar disk masses.

The effects of dust continuum opacity on the protostellar disk kinematics can only be properly taken into account with modeling. Many of the analytic techniques for fitting protostellar disk rotation curves and protostar masses rely on the assumption of compact, optically-thin gas emission that is not affected by continuum opacity (\ct{Tobin+2012,Yen+2013,Ohashi+2014}). The highest-velocity emission, in addition to having the lowest signal-to-noise, is also coming from the smallest radii, where the disk is the most dense. Therefore, this emission would quite likely have its spatial distribution affected by continuum opacity. Indeed, \citet{VantHoff+2018} showed that in L1527 IRS, the emission at the very highest velocities shifts slightly to larger radii, indicating that the higher velocity emission from smaller radii is being attenuated and the assumption of the line velocity directly mapping to a radius of the disk breaks down.

Spectral line modeling is also not without its caveats because it requires a detailed knowledge of the velocity field in both the envelope and disk to properly account for line opacity as a function of velocity (\ct{VantHoff+2018}). Furthermore, the abundance profile of the molecule used for modeling should be well-characterized, otherwise a model cannot accurately account for the spatial variation of the molecule and modeling could represent abundance variations as density structure variations. Therefore, a balanced approach should attempt to employ multiple techniques and molecules when possible to obtain the most realistic results.

Note that the opacity of the dust emission can also pose challenges for the characterization of molecular line emission from protostellar disks. The typically higher mass of protostellar disks as compared to proto-planetary disks (\ct{Tobin+2015b,Tychoniec+2018,Maury+2019,Williams+2019,Tobin+2020}) can result in the disks being optically thick submm/mm wavelengths. This not only prevents tracing all the disk mass from the dust continuum, but the opaque continuum will also absorb molecular line emission, preventing the measurement of molecular gas emission and the characterization of the kinematics.

Furthermore, for high-mass sources, the number of current observational studies targeting high-mass YSOs associated with disks at sufficiently high spatial resolution is still limited. Systematic surveys targeting high-mass star forming regions such as the recent works by \citet{Beuther+2018,Csengeri+2018,Sanna+2018} or on-going ALMA projects such as ALMAGAL, which is targeting more than 1000 high-mass star forming regions in the Galaxy at sub-arcsecond resolution will provide enough high-mass disk candidates to be studied a high spatial resolution. It is also worth noting that most high-mass YSOs are deeply embedded in dust cores optically thick at millimeter and submillimeter wavelengths, thus obscuring the innermost regions where the disk resides. For such objects, observations at longer wavelengths (or shorter frequencies) may be necessary, and for this, instruments such as the SKA and the ng-VLA will complement the current studies conducted with ALMA.

\subsection{Connect theoretical models with observations: synthetic observations}
\label{S.Synthetic}

As both observational and theoretical studies get advanced, our understanding of star formation processes become more detailed and quantitative. It is not trivial, however, to compare sophisticated theoretical models and high-quality observations. While theoretical models provide ``raw" quantities such as the gas density and temperature, what we observe are radiation quantities such as spectral energy distribution (SED), intensity distribution, and channel maps. Also, observations provide only limited information such as position-position intensity distribution and position-position-velocity channel maps with finite resolution. Therefore, to directly compare theoretical models with observations, we need to convert the model data into observables, and simulate observations. Such a procedure, often done as post processing on numerical simulation data, is called synthetic observation, and has been rapidly developing. This technique is useful for interpreting observations but also for predicting observations, estimating required resolutions and sensitivities for instance.

The simplest form of synthetic observations is radiation transfer simulation of dust continuum. This is often done in two steps, first calculating the dust temperature assuming equilibrium between dust emission and irradiation from the central star, and then calculating the intensity distribution from some view points. While our knowledge on the dust opacity is still limited, we can estimate the mass and temperature distribution by comparing actual observations and synthetic observation results. For example, \citet{Maury+2010} performed synthetic observations of numerical simulations with and without magnetic fields to compare with observations of young protoplanetary disks, and concluded that angular momentum in protoplanetary disks must be transported by magnetic fields to explain the small sizes of the observed disks. \citet{Offner+2012} calculated SEDs from protostars forming in turbulent clouds and pointed out that SED is not very accurate indicator of their ages, which questions reliability of the traditional Class-0/I/II categorization. \citet{Tomida+2017} compared an MHD simulation model with Elias 2-27 and showed that the spiral arms in the observed disk are consistent with ones produced by gravitational instability. 

Also, it is possible to include polarization of the dust continuum, which can be produced by various processes including alignment to magnetic fields or scattering by large grains, and infer magnetic field geometry or dust size distribution. This is particularly important because integration of polarized radiation along a line of sight often produces non-trivial structures as it is integration of vector, not scalar. For example, \citet{Hull+2017} compared dust polarization observations of a few star forming clouds in Serpens with simulations of turbulent clouds, and demonstrated that turbulence is dominating the dynamics in the molecular clouds. \citet{Maury+2018} calculated a synthetic polarization map based on an MHD simulation and compared it with B335, a very young Class-0 object with a very compact (or no) disk. They concluded that the B335 system is strongly magnetized, and the small disk size is a consequence of strong magnetic braking. \citet{Kataoka+2015} demonstrated that large grains can produce characteristic patterns in protoplanetary disks by scattering, and proposed a method to constrain the dust size by measuring polarization degrees at different wavelengths. 

Simulations of molecular lines are much more complicated because it involves calculation of chemical abundances and excitation states of each molecule. For now, it is common to adopt simple assumptions such as constant chemical abundances and local thermodynamic equilibrium, but it is possible to include these realistic effects. These synthetic observations are powerful as they can provide richer information about dynamics and chemical reactions. For instance, \citet{Frimann+2016} proposed that the spatial extent of C$^{18}$O can trace the recent burst activity or history of episodic accretion in protostars based on synthetic C$^{18}$O map created from an MHD simulation (see also \ct{Hsieh+2019b}).

For synthetic observations, some publicly available radiation transfer codes are often used, including RADMC-3D (\ct{Dullemond+2014}), Lime (\ct{BrinchHogerheijde2010}), Hyperion (\ct{Robitaille2011}), POLARIS (\ct{Reissl+2016}), TORUS (\ct{Harries+2019}) and so on. Also, to simulate observations with ALMA and other radio interferometers, the standard ALMA analysis software CASA is often used as it contains capabilities to simulate interferometric observations using actual array configurations.

\subsection{Limitations of current numerical simulations}

Despite numerous progress, the numerical modelling of disk formation is currently hampered by several uncertainties that we briefly review here. 

First of all, it has been found that the treatment of the sink particles, used to model the star, has a strong influence on the disk itself, particularly on its mass. For instance, \citet{Machida+2014} explored the influence of various types of criteria to decide how accretion proceeds onto the sink. When using a criterion which relies on the density just outside the sink, they found that the disk is very small or sometimes does not even exist. On the other-hand they infer that when a density is prescribed as a threshold above which accretion occurs,  a disk generally forms but its mass and size vary with the chosen density threshold. Similar conclusion is reached by \citet{Hennebelle+2020} who also pointed out that the density threshold is resolution dependent. That is to say, the same threshold does not has the same influence at different spatial resolution. They interpreted this as due to the fact that the sink particle provides an inner boundary condition for the density. While the density profile is essentially independent of the density threshold applied to the sink, its value determine the largest density value and therefore the mass of the whole disk. \citet{Hennebelle+2020}  speculated that in reality this condition would be determined by the star-disk connection and inferred a simple model for the accretion density threshold that should be used. They indeed found that it is resolution dependent. 
 
Another source of uncertainties comes from the diffusivities that are used to describe the Ohmic, ambipolar diffusion and Hall effects. In particular, as already stressed above, they depend a lot on the grain size distribution \citep{Zhao+2018a} and on the cosmic ray ionization \citep{WursterLi2018}, which are both poorly known.

\subsection{Implications for planet formation}
\label{S.Planet}

Since protoplanetary disks provide the initial and boundary conditions of the planet formation, their structure and evolution have significant impact on the planet formation processes. As the initial and boundary conditions of the planet formation, most studies so far have assumed disk models with simple power-law profiles, including the Minimum Mass Solar Nebula (MMSN; \ct{Weidenschilling1977,Hayashi1981}). While such models worked fine to some extent, they are not sufficient to explain the diversity of the observed exoplanet systems.

The MHD simulations of the star and disk formation tend to form relatively compact but massive disks in the early embedded (Class-0/I to early Class-II) phases, because magnetic fields efficiently transport angular momentum and the accretion rate is high. The structure (e.g. the density and temperature profiles) of such a massive disk is very different from the conventional disk models like MMSN. In such massive disks, it is possible that the dust grains are already grown up significantly since the gas and dust densities are high (\ct{Tsukamoto+2017}). Therefore, the planet formation can be already started even in the early phase of star formation where the star and disk are still growing by accretion. This also means that environmental parameters such as initial turbulence and magnetization can affect resulting planets, which may explain the diversity of exoplanetary systems. In this sense, we need more consistent scenario to consistently integrate star, disk and planet formation processes. At the very least, we need to consider much broader disk models as the initial and boundary conditions of planet formation studies.


\section{Summary \& Outlook}
\label{Chap.Summary}

In this article, we have reviewed the recent progress on  disk and outflow observations and theories for both low and high mass stars. The advancement in high-resolution and high-sensitivity instruments in the past decades have revealed unprecedented details of the structures around young stellar objects. The increasing observational evidence of rotationally supported structures (disks, spirals, multiples) challenges the previously established paradigm of protostellar collapse of magnetized cloud cores, where disk formation is strongly suppressed by magnetic braking. 

After a series of theoretical and numerical exploration in the past few years for low mass case, non-ideal MHD effects with an enhanced level of magnetic diffusivities than previously adopted have shown to be plausible solutions for averting the magnetic braking catastrophe. Non-axisymmetric perturbations of the initial core, such as magnetic field misalignment and turbulence, also promote the initial formation of circumstellar disks.
With the alleviation of the magnetic braking catastrophe, circumstellar disks of 20--30~au commonly form in recent non-ideal MHD simulations, some disks are connected with extended spiral arm structures, when sufficient angular momentum is still left in the inner envelope. For cores with lower magnetization, fragments and clumps often form as the mass accretion rate from the envelope remains high; such a system tends to evolve into binary or multiple stellar systems, which is converging with observations. 
Furthermore, the tens-of-au disks formed around low mass protostars provide reasonable settings for applying the disk wind theory. Particularly, the outflow rotation and the offset of the outflow launching site observed recently are encouraging evidence for existing disk wind theory. 
On the other hand, the well-collimated jet component, revealed in more details by ALMA, may originate from different launching mechanisms.

Disk formation for isolated high mass star formation can be treated as a scaled-up version of the low mass case. However, the strong radiation from the central stellar object may shape the disk evolution and properties differently than the low mass case. Moreover, the dense and turbulent cloud conditions often result in the formation of hierarchical stellar systems with dynamically interacting disks and outflows. These pictures of high mass disk formation requires further observational support. As the high mass sources locate systematically farther away, current interferometry is still limited in resolving and separating the disk and accretion flows around high mass protostars. It is possible that the compact structures currently observed contain rich and complex sub-structures, to be resolved by future generation instruments.

Nevertheless, limitations still exist in both the  current observation and numerical simulation. The optical depth at millimeter and sub-millimeter becomes large enough at the disk scale. Degeneracy still exists between larger grain sizes and larger optical depth in the central compact disk. Therefore, dedicated longer wavelength (e.g., centimeter) observations, as well as kinematics probed by optically-thin tracers (e.g., faint isotopologues) may shed light on this issue.
Furthermore, the advent of new facilities in the infrared range in the current and next years (e.g., MATISSE at the VLTI, METIS at the ELT) will open a new window to study disks in the infrared achieving high angular (sub-arcsecond) resolutions.
Numerically, multi-scale and multi-physics collapse simulations remains challenging. 
For constructing a consistent picture of star and disk formation, connections with the larger molecular cloud environment may become pivotal in future explorations.

\begin{acknowledgements}
B.Z., K.T., P.H., R.K., A.R., A.B., M.P., and Y.-N.L. thank the staff of ISSI for their generous
hospitality and creating the fruitful cooperation. 
A.S.-M. acknowledges support from the Collaborative Research Centre 956 (subproject A6), funded by the Deutsche Forschungsgemeinschaft (DG) - project 184018867.
A.L.R acknowledges support from NASA through Einstein Postdoctoral Fellowship grant No. PF7-180166 awarded by the Chandra X-ray Center, which is operated by the Smithsonian Astrophysical Observatory for NASA under contract NAS8- 03060.
M.P. acknowledges funding from the INAF PRIN-SKA 2017 program 1.05.01.88.04.
A.M. acknowledges support from European Research Council (ERC) under the European Union Horizon 2020 research and innovation program (MagneticYSOs project grant agreement N.679937). 
The National Radio Astronomy Observatory is a facility of the National Science Foundation operated under cooperative agreement by Associated Universities, Inc.
\end{acknowledgements}


\bibliographystyle{spbasic}      
\bibliography{bibliography}   

%



\end{document}